\documentclass[12pt]{iopart}
\usepackage{graphicx}
\usepackage{xcolor}

\usepackage{amssymb}
\usepackage{grffile} 
\usepackage[inline]{enumitem}

\usepackage[numbers,sort&compress,square]{natbib}

\usepackage{hyperref}
\hypersetup{
    colorlinks=true,
    linkcolor=blue,
    filecolor=magenta,      
    citecolor=magenta, 
    urlcolor=cyan,
    }

\makeatletter
\long\def\@makefntext#1{\parindent 1em\noindent 
 \makebox[1em][l]{\footnotesize\rm$\m@th{^\arabic{footnote}}$}%
 \footnotesize\rm #1}
\def\@makefnmark{\hbox{$^{\arabic{footnote}}\m@th$}}
\def\@thefnmark{\arabic{footnote}}
\makeatother

\newcommand{\signal}{{\sf signal}}
\newcommand{\glitch}{{\sf glitch}}
\newcommand{\noise}{{\sf noise}}

\begin{document}

\title[Neural network time-series classifiers for GW searches in single-detector periods]{Neural network time-series classifiers for gravitational-wave searches in single-detector periods}
\author{A.~Trovato}
\address{Dipartimento di Fisica, Universit\`a di Trieste, I-34127 Trieste, Italy\\
INFN, Sezione di Trieste, I-34127 Trieste, Italy}
\ead{agata.trovato@units.it}
\author{E.~Chassande-Mottin}
\address{Universit\'e Paris Cit\'e, CNRS, Astroparticule et Cosmologie, F-75013 Paris, France}
\author{M.~Bejger}
\address{INFN, Sezione di Ferrara, I-44122 Ferrara, Italy\\
Nicolaus Copernicus Astronomical Center, Polish Academy of Sciences, ul. Bartycka 18, 00-716 Warsaw, Poland}
\author{R.~Flamary}
\address{Ecole polytechnique, IP Paris, CMAP, F-91120 Palaiseau, France}
\author{N.~Courty}
\address{Université Bretagne Sud, CNRS IRISA, F-35042 Rennes, France}

\vspace{10pt}

\begin{abstract}
The search for gravitational-wave signals is limited by non-Gaussian transient noises that mimic astrophysical signals. Temporal coincidence between two or more detectors is used to mitigate contamination by these instrumental glitches. However, when a single detector is in operation, coincidence is impossible, and other strategies have to be used. We explore the possibility of using neural network classifiers and present the results obtained with three types of architectures: convolutional neural network, temporal convolutional network, and inception time. The last two architectures are specifically designed to process time-series data. The classifiers are trained on a month of data from the LIGO Livingston detector during the first observing run (O1) to identify data segments that include the signature of a binary black hole merger. Their performances are assessed and compared. 
We then apply trained classifiers to the remaining three months of O1 data, focusing specifically on single-detector times. The most promising candidate from our search is 2016-01-04 12:24:17 UTC.
Although we are not able to constrain the significance of this event to the level conventionally followed in gravitational-wave searches, we show that the signal is compatible with the merger of two black holes with masses $m_1 = 50.7^{+10.4}_{-8.9}\,M_{\odot}$ and $m_2 = 24.4^{+20.2}_{-9.3}\,M_{\odot}$ at the luminosity distance of $d_L = 564^{+812}_{-338}\,\mathrm{Mpc}$.
\end{abstract}

\section{Introduction}
The breakthrough discovery of gravitational waves (GW) on September 14, 2015 \cite{GW150914}, announced by the LIGO Scientific Collaboration \cite{aasi15:_advan_ligo} and the Virgo Collaboration \cite{acernese15:_advan_virgo}, opened the era of the GW astronomy. The detection happened during the first observing run (O1) of the LIGO detector.
With the subsequent observing runs, O2 and O3, performed jointly with Virgo, the list of detected GW signals has grown to 90 events. While the detected sources are mainly associated with the merger of binary black holes (BBH), they also include binary systems with neutron stars \cite{2017PhRvL.119p1101A,2020arXiv200101761T,Abbott_2020,2021ApJ...915L...5A}. These detections are collected and characterized in the GW transient catalogs GWTC \cite{2019PhRvX...9c1040A, 2021PhRvX..11b1053A, 2021arXiv210801045T, 2021arXiv211103606T}.
On May 2023 the fourth observing run (O4) started with an increasing detector sensitivity and consequently an enhanced expected rate of detections.

GW transient signals are detected in the data by a variety of data analysis pipelines, see e.g. \cite{2021arXiv211103606T} for a recent review. In particular, matched filtering \cite{creighton11:_gravit_wave_physic_astron} is a prominent technique to search for signals when an accurate waveform model is available, as in the case of compact star binary mergers. Algorithmically, this consists in correlating the data with a large set of template waveform models (the ``template bank'') that are representative of all the morphologies the expected signal can possibly take.

To make robust detection statements, those pipelines have to address a major difficulty: the presence in the data of short-duration noise artefacts, often called ``instrumental glitches'' \cite{LIGO:2021ppb,Virgo:2022fxr}, that can mimic the GW signal \cite{2023arXiv230409977A,2023PhRvD.107b4030C}. A very powerful tool to discriminate the signal from noise glitches is time coincidence across two or more separate detectors (see \cite{Dhurandhar:2010tc} for a discussion on multi-detector noise rejection techniques).

Obviously, coincidence cannot be used during periods when only one detector operates. During the O1 and O2 observing runs, single-detector periods amount to about 30\% of the observation time \cite{O1status, O2status}. During O3, thanks to a more stable and reliable operation, this fraction was reduced to about 15\% in O3a \cite{O3astatus} and 11\% in O3b \cite{O3bstatus} (the first and second six months parts of O3). In total, more than five months of observing time fall in this category, so far. The O4 science run initiated recently may also have long periods of single detector times.

The lack of coincidence results in difficulties to disentangle the signal from glitches and to measure the statistical significance of a trigger to high confidence levels. Several studies investigate ways to resolve these difficulties. Two methods \cite{PhysRevD.95.042001, PhysRevD.98.024050} that allow the identification of gravitational-wave candidates in single-detector data have been employed in production in the context of low-latency gravitational-wave searches \cite{Abbott_2019c}, enabling the initial identification of GW170817 and GW190425. Similarly, Ref. \cite{Callister:2017urp} introduces a framework for assigning significance to single-detector gravitational-wave events by leveraging the measured rate of binary black hole mergers. More recently, ref.~\cite{Sachdev:2019vvd} studies the possibility to extend the multi-variate likelihood-ratio statistics used by the {\tt GstLAL} pipeline to generate single-detector events. The likelihood estimation has been recently updated in view of the O4 run \cite{Tsukada:2023} and one of the improvements is the addition of a tuneable penalty in case of single detectors candidates to down weight their significance \cite{Ewing:2023}. To extrapolate the significance measure of single-detector triggers produced by the {\tt PyCBC} pipeline \cite{pycbc}, a method proposed in \cite{Davies:2022thw} allows to recover loud signals in single-detector data. In both cases, it is shown that the search sensitivity is significantly reduced compared to multi-detector searches.

Despite those developments, single-detector periods have received less attention than the rest of the observations and are covered in a few studies. Following a ``multi-messenger'' approach, several works looked for coincidences between data from a solitary gravitational-wave detector with gamma-ray observations from the Fermi Gamma-ray Burst Monitor \cite{GW-GRB, GW-BNS151030, Stachie_2020}. Three searches for binary mergers in single-detector periods relied on gravitational-wave data only. Ref. \cite{GW-BNS} presents a search which specifically targets a narrow range of low masses motivated by the population of known double neutron-star binaries. Two contributions present the results of searches for binary mergers over the entire range from 1 to 100 $M_{\odot}$ for the component masses, for the observing runs O1 and O2 \cite{Nitz:2020naa} and for O3 \cite{Davies:2022thw}. The former finds two candidate events observed in single detector periods: 2015-12-25 04:11:44~UTC with the LIGO Hanford detector and 2016-01-04 12:24:17~UTC with the LIGO Livingston detector. The first candidate event has a low significance with a probability of astrophysical origin \cite{LIGOScientific:2016kwr} $p_{\mathrm{astro}} = 0.12$, while the second has a larger significance $p_{\mathrm{astro}} = 0.47$. However, for this event, an excess power observed in the residual after subtraction of the best-fit waveform from the data suggests this event may not be of astrophysical origin, and is thus discarded.

Glitches of different types vary widely in duration, frequency range and morphology. It is difficult to construct a statistical model able to capture the overall complexity of the glitch populations. Their complex and time-evolving nature makes glitch identification and rejection a good problem and a use case for machine learning (ML). In principle, this approach allows to train a classifier able to distinguish between different types of input (glitches versus real GW signal in our case), and thus to learn a possibly very complex and high-dimensional statistical model from a set of examples.

As in many scientific fields, the use of ML has recently gained in popularity in the context of GW astronomy. There is a fairly large body of works pertaining to various aspects ranging from denoising, glitch classification and cancellation, waveform modelling, searches for GW signals, astrophysical parameter estimation, population studies (see e.g. \cite{Cuoco2020, Huerta2020} for recent reviews).

In the context of GW signal searches, convolutional neural networks (CNN) \cite{GoodBengCour16} have been investigated to detect BBH signals for both single- and multi-detector cases \cite{PhysRevD.97.044039, GEORGE201864, PhysRevLett.120.141103, KRASTEV2020135330, PhysRevD.100.063015, 2022PhRvD.105d3002S}. The primary motivation put forward in those contributions is the computational gains expected from the use of CNNs compared to matched filtering techniques.

So far a large fraction of those investigations use simulated Gaussian noise \cite{PhysRevD.97.044039, PhysRevLett.120.141103, KRASTEV2020135330, 2022PhRvD.105d3002S}. In this case, it is not possible to learn the non-Gaussian component of the instrumental noise. Few studies use real GW data including glitches \cite{GEORGE201864, PhysRevD.100.063015}. The classifiers obtained in those contributions are limited to false positive probability (i.e., noise or glitches classified as signal) of about 1\%. This corresponds to a false alarm rate of once every 40 minutes, which is not sufficient in practice. A recent review \cite{Schafer:2022dxv} compares different approaches on a mock data challenge.

The purpose of this study is to enhance the ability of neural network based searches to reject noise artifacts and improve their sensitivity, with a particular focus on analyzing data from a single detector. The goal is to achieve a false alarm rate similar to that of current online searches performed by the LIGO-Virgo-KAGRA collaboration (LVK), i.e. two false alarms per day \cite{ligo_virgo_kagra_public_alert_user_guide}. We explore various network architectures, particularly those designed for time series classification \cite{TCN, IT}.

We trained and tested neural network classifiers using a dataset produced from one month of O1 data collected by the LIGO Livingston detector, during which no GW signals were detected using the matched filtering based searches. 

Section \ref{sec:setup} provides details on how the training and testing sets are generated, while Sec.~\ref{sec:models} describes the structure of the various neural network classifiers being considered. The performance and efficiency of the classifiers are assessed using testing data, and the results are presented in Sec.~\ref{sec:results}. We applied these classifiers to the remaining three months of O1 data, including segments associated with the three GW events detected during O1. Sec.~\ref{sec:remaining} summarizes the results of this analysis. We checked the classifiers' response obtained with the known detected events during O1. A particular focus is then given to the single-detector times. Interestingly, we found that only one data segment was classified as ``signal" by the three classifiers we considered. This event coincides with the single-detector event found by \cite{Nitz:2020naa} in the LIGO Livingston data, as mentioned above, and was downgraded by the same study as a noise artifact. Following the additional checks we conducted on this event, we arrived at a different conclusion as they confirmed its compatibility with an astrophysical origin. Finally, Sec.~\ref{sec:conclusion} concludes on the applicability of the proposed methodology.

\section{Generation of datasets for training and testing}
\label{sec:setup}

The typical approach for applying ML methods to GW detection is to treat it as a classification problem, see e.g., \cite{PhysRevD.97.044039, GEORGE201864, PhysRevLett.120.141103, KRASTEV2020135330, 2022PhRvD.105d3002S}. In this approach, we aim to determine whether a given segment of GW strain data of fixed duration contains an astrophysical signal or not. This problem can be solved by developing an ML-based classifier that is trained using example data. We produce training data labeled as follows:

\begin{itemize}
\item {\noise}: the data are compatible with stationary background noise, i.e., are free of transient instrumental artifacts (glitches) or known GW events,
\item {\glitch}: the data include one or several transient instrumental artifacts (glitches),
\item {\signal}: the data include a (simulated) astrophysical signal, added to the stationary background noise.
\end{itemize}

This three-class approach differs from other contributions in the literature, which consider only two classes. The presence of glitches is known to significantly alter the statistical distribution of the data. By assigning a specific label to data segments containing glitches, the idea is that this may aid the classifier in achieving improved performance. Furthermore, the relative significance assigned to each class could offer valuable information when evaluating the contents of a given segment.

Training and testing data are extracted from the dataset of the observing run O1, which was publicly released via the Gravitational Wave Open Science Center (GWOSC) \cite{gwosc}. Specifically, we utilize the data from the LIGO Livingston detector spanning one month between November 25, 2015 (GPS time 1132444817) and December 25, 2015 (GPS time 1135036817). Throughout this duration, no GW signals were detected by the standard search pipelines.

In this period the available L1 data amounts in total to about 13.3 days (1,147,457 s), of which 3.6 days (312,284 s) were in single-detector time, i.e. 27\% of the time.

The raw data are sampled at 16 kHz. We have downsampled the data to 2048 Hz,\footnote{The method {\tt signal.decimate} of the software package  {\tt Scipy} \cite{2020SciPy-NMeth} is used to downsample.} bandpass-filtered between 20 Hz and 1 kHz and whitened by applying the inverse amplitude spectral density (ASD) in the frequency domain.\footnote{For the preparation of the training and testing data, we acknowledge the use of the following software packages: {\tt GWpy} \cite{gwpy}, {\tt PyCBC} \cite{pycbc} and {\tt LALSuite} \cite{lalsuite}.} The ASD is estimated over stretches of variable length, depending on the duration of uninterrupted data-taking periods (minimum duration is 37 s and maximum is 100,573 s). The data are divided into one-second non-overlapping segments.

\begin{figure}
  \centering
  \includegraphics[width=0.95\linewidth, trim={3cm 0cm 4cm 0cm}]{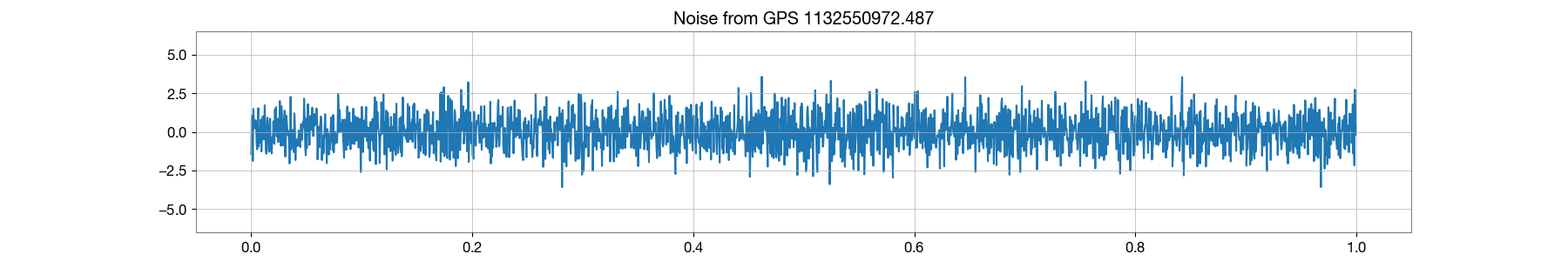}
  \includegraphics[width=0.95\linewidth, trim={3cm 0cm 4cm 0cm}]{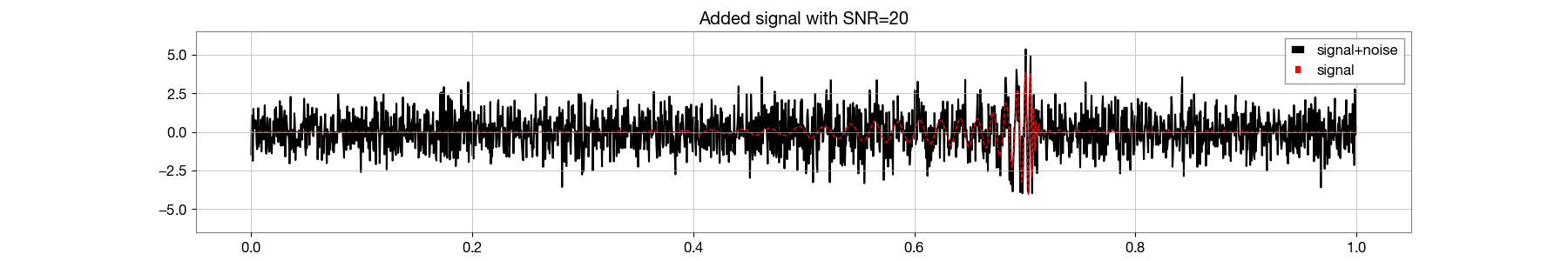}
  \includegraphics[width=0.95\linewidth, trim={3cm 0cm 4cm 0cm}]{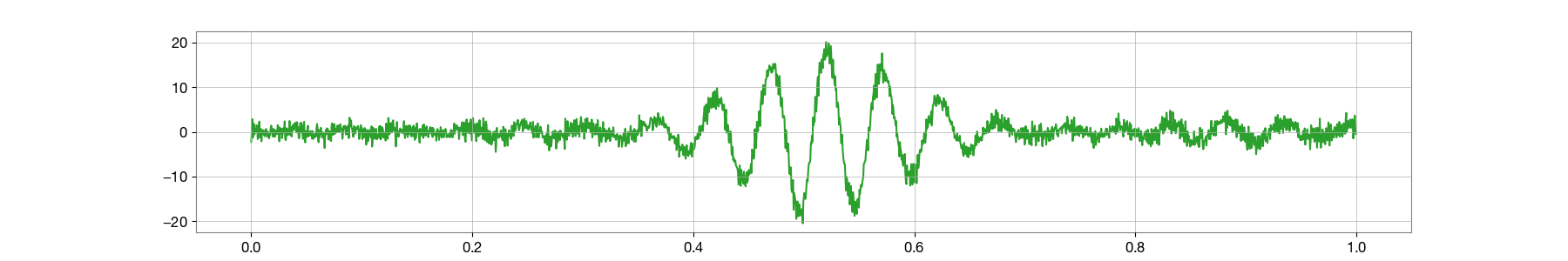}  
  \caption{Instances of the classes {\noise} (blue), {\signal} (black) and {\glitch} (green). Top ({\noise}): one-second data segment recorded by LIGO Livingston at the GPS time 1132550972.487. Middle ({\signal}): a simulated BBH waveform with SNR of 20 (dashed red line) is injected in the previous timeseries. Bottom ({\glitch}): Data recorded at the GPS time 1132580628.41 which contains a low-frequency transient instrumental artifact.}
  \label{fig:3classes}
\end{figure}

The data are distributed into the three classes introduced above as explained in the next sections. Representative instances of the three classes are shown in Fig.~\ref{fig:3classes}.

\subsection{The {\noise} class}
\label{sec:noise}

The {\noise} class corresponds to segments that are free of known GW signals, glitches (see next section) or hardware injections.\footnote{During O1, hardware injections, which are simulated signals created by manipulating mirrors in the arms of the interferometers, were added to the LIGO detectors for testing and calibration. See \url{https://www.gw-openscience.org/o1_inj}} All segments in the dataset passed the first criterion, as no GW signals were confidently detected by standard pipeline over the selected period.\footnote{This implies that the {\noise} label is essentially determined by the sensitivity limit of the matched-filtering based searches.} Overall, there is a total of 750,000 {\noise} samples in the one-month O1 dataset.

\subsection{The {\glitch} class}
\label{sec:glitch}

A database of glitches is created using two different sources: the unmodeled transient search \emph{coherent WaveBurst} ({\tt cWB}) \cite{cWB, klimenko_sergey_2021_4419902} and the citizen science project {\tt Gravity Spy} \cite{BAHAADINI2018172}.

The cWB pipeline is an open-source software package designed to search for a wide range of GW transients without prior knowledge of the signal waveform. To evaluate the analysis background, cWB uses a resampling technique \cite{cWB} that involves applying non-physical time shifts to the data before analysis. Loud, i.e., high signal-to-noise ratio (SNR), background triggers resulting from this procedure are good candidates for glitches. The loudest triggers in LIGO Livingston with an SNR higher than 5.8 were selected (258,480 glitches). This list was complemented with the {\tt Gravity Spy} database (13,144 glitches). The timestamps and duration of the identified glitches from these two sources are collected in a single list, which is then used to label the one-second data segments from the O1 observing run. If the glitch duration is shorter than 1 second, the associated segment is labeled as a {\glitch}. Note that the glitch has a random position within the one-second window. If the glitch duration is longer than 1 second, all segments that overlap with that glitch duration are labeled as {\glitch}. Only the glitches whose time belongs to the data segments available on GWOSC are considered. In many cases, the glitches are closer in time than one second, so multiple glitches can fall in the same one-second segment.

From the one-month O1 data, a total of $150,000$ segments receive the {\glitch} label.

\subsection{The {\signal} class}
\label{sec:signal}

The samples from the {\signal} class are produced by adding simulated GW signals from BBH systems to the one-month O1 data in periods without known GW signals or hardware injections. 
For the training set, the data segments used to generate samples of the {\signal} class are not utilized for the {\noise} class nor the {\glitch} class, while for the testing set, the same data segments are used for both the {\noise} and {\signal} classes. 
To generate the astrophysical signals, the waveform model \texttt{SEOBNRv4} \cite{2017PhRvD..95d4028B} is employed, with a lower frequency cutoff of 30 Hz. The simulated signals are sampled, whitened, and band-pass filtered in the same manner as the data segments.

The masses of the binary BH used for generating the simulated signals in the class {\signal} are chosen to ensure that they fall within the mass range observed by the LVK and that the signals are short enough to be contained within the one-second data segments. Specifically, the component masses $m_1$ and $m_2$ are chosen randomly, with the constraint that $m_1 > m_2 \geq 10 M_{\odot}$ and the total mass $M = m_1 + m_2$ is uniformly distributed in $33 M_{\odot} \leq M \leq 60 M_{\odot}$. We consider non-spinning BH, so the dimensionless spin magnitudes $\chi_1$ and $\chi_2$ are set to 0. The phase at coalescence and the polarization angle are drawn uniformly in $(0, 2\pi)$, and the inclination angle in $(0, \pi)$. Since the focus is on a single detector, the right ascension and declination are not particularly important and are thus fixed to zero. 

The amplitude of the added signals is computed such that the corresponding {\em optimal} SNR ${\rho_{opt}}$ is uniformly distributed between 8 and 20. Following \cite{maggiore}, it is defined as 
\begin{equation}
{\rho^{2}_{opt}} = 4\int\nolimits_0^\infty {{{\vert \tilde h(f){\vert ^2}} \over {{S_n}(f)}}{\rm{d}}f}, 
\label{eq:snr_opt}
\end{equation} 
where $\tilde h(f)$ denotes the Fourier transform of the template $h(t)$ and $S_n(f)$ is the power spectral density of the detector noise. To generate the signals, a fiducial luminosity distance $d_{L}$ of 100 Mpc is initially chosen, and then scaled to obtain the desired ${\rho_{opt}}$. The final values of $d_{L}$ range from 1 to 1300 Mpc approximately.

The simulated signals are added at a random position within the segment while ensuring the chirping part of the signal is completely contained in the segment. The final part of the signal is randomly shifted between -0.25 s and 0.3 s with respect to the center of the one-second segment. A total of 750,000 {\signal} samples are generated.

\vskip 2mm
Overall, the training set consists of $250,000$ segments for the {\noise} class, the same number for the {\signal} class, and $70,000$ for the {\glitch} class. A 20\% fraction of the training set is allocated for validation. The testing set, used to evaluate the classifier, comprises $500,000$ samples for both the {\noise} and {\signal} classes, and $80,000$ for the {\glitch} class. This ensures sufficient statistical data for characterizing the classifier's performance. In total, the training and testing datasets comprise 1,650,000 one-second segments, with 45\% for the {\noise} class, 45\% for the {\signal} class, and 10\% for the {\glitch} class. This amounts to a storage space of 26 gigabytes. Out of the total number of segments, 28\% is utilized for training, 7\% for validation, and 65\% for testing.

\section{Classifier architectures}
\label{sec:models}

This section discusses the type of neural network architectures considered in this study. Similarly to other works \cite{PhysRevD.97.044039, GEORGE201864, PhysRevLett.120.141103, KRASTEV2020135330, PhysRevD.100.063015, 2022PhRvD.105d3002S}, the classifier is directly fed by the one-second segment of strain time series, so a vector size of 2048. We experiment\footnote{Implementations are based on the {\tt TensorFlow} library \cite{tensorflow2015-whitepaper} with the {\tt Keras} API \cite{chollet2015keras}.} with three different network architectures, namely the CNN, as well as two other architectures specialised for time-series classification: Temporal Convolutional Network (TCN) \cite{TCN} and Inception Time (IT) \cite{IT}. 
The last two, to our knowledge, have never been tested with this type of problem. The architectures are described in more detail in the following subsections. 
The model hyperparameters provided below have been tuned after a coarse exploration of the parameter space.

\subsection{Convolutional Neural Network (CNN)}

CNNs were first introduced for image classification \cite{GoodBengCour16}. They are now used for a wide variety of tasks, including the detection of GW signals \cite{PhysRevD.97.044039, GEORGE201864, PhysRevLett.120.141103, KRASTEV2020135330, PhysRevD.100.063015, 2022PhRvD.105d3002S}. In this study, we tested a range of CNNs similar to those considered in previous works.

We limited ourselves to shallow networks with five layers, four convolutional layers, and one final fully connected embedding layer. For simplicity, we only report here on the best-performing CNN, whose structure is detailed in Table~\ref{tab:CNN}.

The convolutional layers are defined by the number of output filters, the length of the 1D convolution window (kernel size), the stride length of the convolution, and the activation function. The dense layer only requires the definition of the activation function. The input of inner convolutional layers is downsampled with a max pooling operation over a window size indicated in the table. The output of convolutional layers is processed by a dropout layer that randomly sets the input units to 0 with the frequency rate specified in the table. A global average pooling, followed by a dropout with a rate of 10\%, is applied to the output of the last convolutional layer.

\begin{table}[ht]
\centering
\caption{Structure of the CNN considered in this study. The type of the layer is either convolutional (Conv) or fully connected (Dense). The activation function is either the rectified linear unit ({\tt relu}) or the {\tt softmax} function \cite{GoodBengCour16}.}
\vspace{10pt}
\label{tab:CNN}
{\begin{tabular}{ | l | c | c | c | c | c |}
\hline
\textbf{Layer number}	&	1	&	2		&	3		&	4	&	5\\
\hline
\textbf{Type}			&	Conv	&	Conv		&	Conv		&	Conv	&	Dense\\
\hline
\textbf{Number of filters}			&	256	&	128		&	64		&	64	&	-\\
\hline
\textbf{Kernel size}		&	16	&	8		&	8		&	4	&	-\\
\hline
\textbf{Stride length}		&	4	&	2		&	2		&	1	&	-\\
\hline
\textbf{Activation function}		&	\texttt{relu}	&	\texttt{relu}		&	\texttt{relu}		&	\texttt{relu}	&	\texttt{softmax}\\
\hline
\textbf{Dropout rate}		&	0.5	&	0.5		&	0.25		&	0.25	&	-\\
\hline
\textbf{Max pooling}		&	4	&	4		&	2		&	2	&	-\\
\hline
\end{tabular}
}
\end{table}  

\subsection{Temporal Convolutional Network (TCN)}

TCN \cite{TCN,KerasTCN} is a neural network architecture specifically developed for sequence modeling problems. TCN has been shown to outperform generic state-of-the-art architectures over a diverse range of tasks and datasets. The TCN architecture is based on causal convolutions, where an output at time $t$ is only convolved with past inputs from the previous layer. This allows the network to collect information from further in the past, using a combination of deeper networks (augmented with residual layers) and dilated convolutions.

In this study, we have tuned the hyperparameters of the TCN model to find a compromise between the best performance and a reasonable training time. We ended up using a network with a TCN layer consisting of $N=6$ dilated convolutional layers with 32 filters, a kernel size of $k=16$, default values of dilation factors $d_{k=1 \ldots 6}=(1, 2, 4, 8, 16, 32)$ for the 6 convolutional layers, and a dropout rate of 0.1. The output of the TCN layer goes into a final dropout layer with a rate of 0.5, and a dense embedding layer closes the model.

A key parameter that governs the training efficiency is the receptive field, which is the size of the region in the input data that produces a given feature in the output. The receptive field of the TCN can be expressed as $R=1 \,+ \,2 \,(k-1)\, d_\mathrm{tot}$ where $d_\mathrm{tot}=\sum d_k$ \cite{TCN}. With the above configuration, we have $R \approx 1900$. The data used in this work have a sampling rate of 2048 Hz so each segment of data has 2048 points. The training with TCN is effective when $R$ is much larger than the length of the input sequence \cite{TCN}. To satisfy this constraint, only for this model, it is necessary to downsample the input data to 1024 Hz, therefore producing an input vector of size 1024.

\subsection{Inception Time (IT)}

IT \cite{IT} is a deep network ensemble designed specifically for time series classification. It leverages the concept of residual networks and incorporates Inception modules \cite{2014arXiv1409.4842S}. In a nutshell, the Inception module first produces a one-dimensional summary of the input multivariate time series (this is the ``bottleneck'' layer), and then convolves this summary through multiple filters of different lengths, leading to a multivariate output that provides inherently multi-resolution features. The module output is finally reduced by max pooling (pool size of 4) before passing to the next module.

The IT architecture is composed of five ResNet networks with a sequence of depth $d$ Inception modules, with two residual blocks. The outputs of the five models are combined through a global average pooling and a final softmax layer, used to produce the classification probabilities for the different classes. In this study we have used the standard implementation of IT provided by the authors \cite{IT} with networks of depth $d=10$, each with a bottleneck size of 32 processed through 32 filters with kernel sizes 20, 40 and 80.

\subsection{Training process}

The three classifiers are optimized using the training set described in Sec.~\ref{sec:setup} to minimize the categorical cross-entropy loss function. The default implementations of the Adam optimizer are utilized, with a batch size of 24 \cite{tensorflow2015-whitepaper}. The training procedure is repeated 10 times with different (random) initializations of the model weights and dropouts, and the instance exhibiting the best Receiver Operating Characteristic (ROC) curve on the testing dataset (as explained in Sec.~\ref{sec:results}) is chosen. Note that this evaluation cannot be done with the validation dataset, as it does not provide enough statistics to compute the ROC in the relevant regime of low false alarm rates.

Throughout the training process, the model's area under the ROC curve \cite{FAWCETT2006861} is evaluated on the validation data, and the model with the highest value is ultimately selected. The CNN, TCN, and IT models are trained for 50, 150, and 20 epochs, respectively. The best models are obtained at the 24th epoch for CNN, the 34th epoch for TCN, and the 5th epoch for IT\footnote{After the 5th epoch, the IT model displayed signs of overfitting.}. On the Tesla K40d GPU we used, the training times per epoch were 220 seconds for CNN, 1000 seconds for TCN, and 3320 seconds for IT.

\subsection{Decision statistic}
\label{sec:softmax}

The final objective is to detect with high confidence the segments with a true astrophysical signal, i.e., to classify them as {\signal} and to reject the other segments as {\noise} or {\glitch}.\footnote{The two classes {\noise} and {\glitch} will be later merged \textit{a posteriori} into a single class associated with the absence of an astrophysical signal.} We aim to constrain false alarms to a rate of two per day (similar to the current online search pipelines). This implies that we should reject all but one {\noise} or {\glitch} segment from the testing set in $1.7 \times 10^5$ trials.

The classifiers output the probability of class membership for each of the classes, that is three numbers between 0 and 1, summing to 1. The final detection is performed by applying a threshold to the membership probability $P_s$ assigned to the {\signal} class, which thus defines our \textit{decision statistic}. The class membership probability is computed by the softmax activation function applied to the raw output (the ``logits tensor'') of the fully connected embedding layer which concludes the classifiers. Because of the high-confidence level required, this threshold is very close to 1, thus requiring attention to the numerical precision for the evaluation of the membership probability (This issue related to the precision of floating-point arithmetics was already noted in \cite{2022PhRvD.105d3002S}). This has consequences on the way the classification loss is computed from the membership probability at the training stage. We found that the categorical cross-entropy loss should be directly computed from the logits tensor rather than from the class membership probability after the softmax transformation.

This numerical precision issue has an impact on the performance of both the TCN and IT classifiers. The right panel of Fig.~\ref{fig:roc} provides an illustration for IT. 
This plot compares the ROC curves based on the detection statistic $P_s$ ( see Sec.~\ref{sec:results} for the details on how the curves are computed) obtained with IT when the categorical cross-entropy loss is calculated from the logits tensor (green) and when it is calculated from the class membership probability after the softmax transformation (red). 
The shaded area represents the range between the best and worst models among the 10 instances computed at training. When softmax is used the uncertainty in the performance is larger (the shaded area is wider) and the classification efficiency reached at small false alarm rates is lower. 

\section{Classifier evaluation with the testing data}
\label{sec:results}

\begin{figure}
  \centering
  \includegraphics[width=\textwidth]{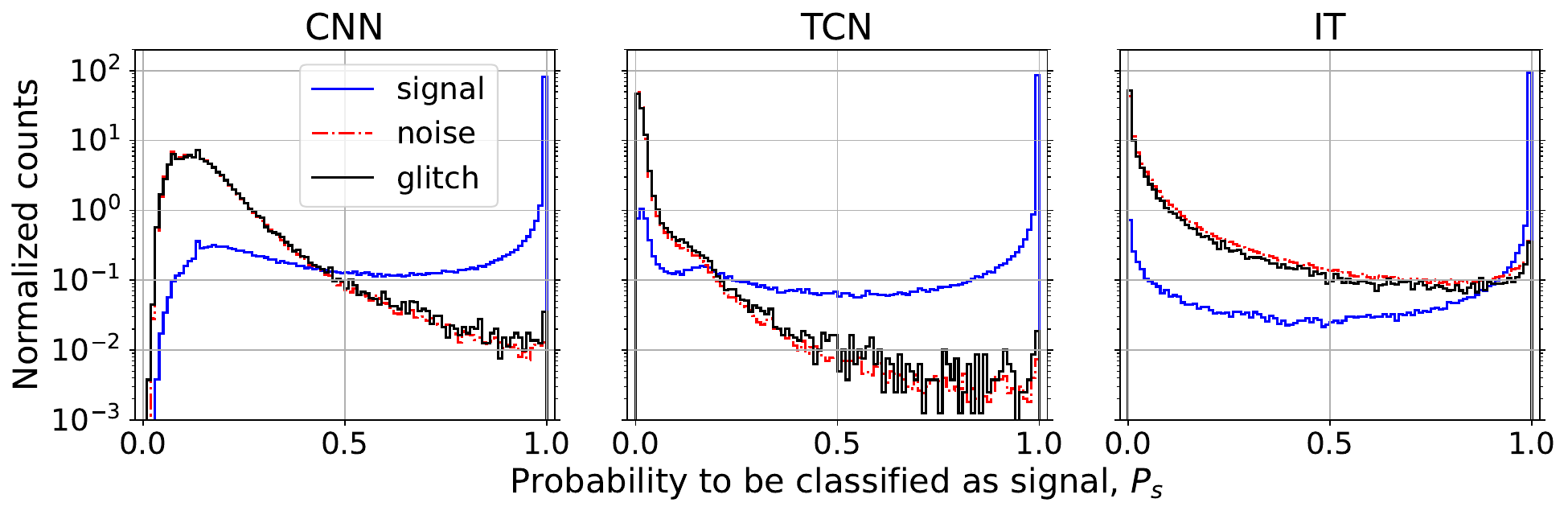}
  \caption{Distributions of $P_s$ (the class membership probability assigned to the {\signal} class), conditioned on the class of the input segment from the testing set: {\signal} (blue), {\noise} (dot-dashed red), or {\glitch} (black). These distributions were computed for the CNN (left), TCN (middle), and IT (right) architectures. The histograms are normalized to have a unit sum. The classifiers do not distinguish between samples from the {\noise} and {\glitch} classes, thus resulting in practically identical probability distributions (see Sec.~\ref{sec:results} for a discussion on this point).}
    \label{fig:probab}
\end{figure}

This section describes the results obtained with the three classifiers presented above applied to the testing set.

The classifiers all exhibit poor separation power between the {\noise} and {\glitch} classes. This can be attributed to several factors, including the absence of a distinct boundary between the two classes (potentially due to contamination and mislabeling), the considerable variation in glitch morphology, and the relative class imbalance with 3.5 times the {\glitch} class being unrepresented by a factor of $3.5$ compared to the {\noise} class in the training set. The initial assumption that a three-class division would enhance classification performance turned out to be incorrect, at least with this dataset. Consequently, we proceed by combining the {\noise} and {\glitch} classes into a single class representing the absence of an astrophysical signal.

\subsection{Noise rejection}
\label{sec:noise_reject}

We first assess the noise rejection capabilities of the classifiers. Fig.~\ref{fig:probab} compares the distributions of the decision statistic $P_s$ (the membership probability assigned to the {\signal} class) when the input segment belongs to each of the three classes. The $P_s$ distributions obtained with samples from the {\noise} or {\glitch} classes have very similar shapes, reflecting the intrinsic similarity of those two classes (see above). The best classifier is the one that provides the greatest contrast between the distributions of the $P_s$ statistic obtained in the presence of a signal (blue) versus noise or glitch (red dot-dashed and black).

The distributions obtained for the {\noise} or {\glitch} classes exhibit maxima at zero for the TCN and IT classifiers, while the maximum is shifted to around 0.1 for CNN. Moving from the peak to higher values, the distribution shows a monotonic decay for CNN and TCN. However, for IT, the distribution initially decreases and then slightly increases near $P_s=1$. The TCN classifier appears to reach the lowest background $\lesssim 10^{-2}$ in normalized count units.

Since our objective is to achieve high-confidence classification, we are primarily interested in the immediate vicinity of $P_s = 1$. This motivates to reparameterize the $P_s$ statistic as $\lambda := -\log_{10}(1-P_s)$. While $P_s$ ranges from 0 to 1, $\lambda$ can theoretically take values across the entire real line. However, our main focus lies in the range $\lambda \gtrsim 7$. The most stringent criterion is to require $P_s = 1$ at machine precision, which corresponds to $\lambda = \infty$. The number of {\noise} and {\glitch} samples in the testing set that satisfy this selection criterion is 0, 1, and 2 for the CNN, TCN, and IT classifiers, respectively. 
Such rejection power (between 0 and 2 false alarms in $5.8 \times 10^5$ trials) is in agreement with the false-alarm rate targeted initially.

\subsection{Signal extraction}
\label{sec:signal_extract}

\begin{figure}
  \centering
  \includegraphics[width=.9\textwidth]{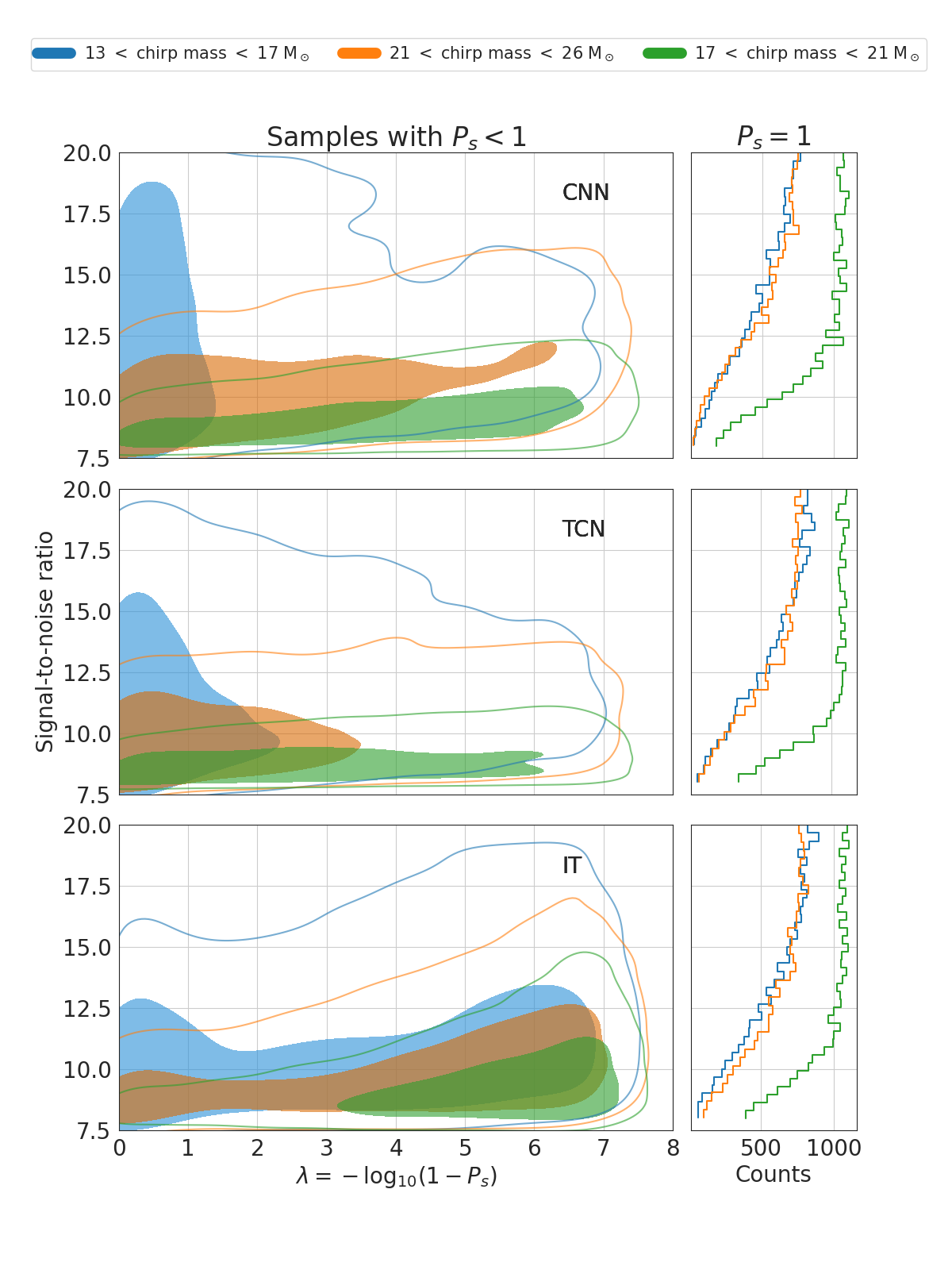}
  \caption{Distribution of the statistic $\lambda := -\log_{10}(1-P_s)$ obtained with testing samples from the {\signal} class and computed for the three considered classifiers: CNN (top), TCN (center) and IT (bottom). The column on the left shows a kernel density estimate of the $\lambda$ distribution for the samples with $P_s < 1$, thus leading to a finite value for $\lambda$. The shaded area is the 50\% containment region, and the line is the 90\% containment region. Those distributions are shown versus the SNR of the injected GW signal and computed separately for three ranges of chirp mass. The column on the right shows a histogram for the samples with $P_s = 1$. The signal samples detectable with high confidence fall in the range of large $\lambda \gtrsim 7$ (i.e, $P_s$ values very close or equal to 1).} 
  \label{fig:probab_vs_snr}
\end{figure}

We proceed to assess the classifiers' ability to extract signals. Fig.~\ref{fig:probab} illustrates the distributions of the decision statistic $P_s$ when the input segment belongs to the {\signal} class, represented in blue. As anticipated, all distributions exhibit a peak at $P_s=1$. However, the peak appears narrower for the IT classifier. To focus on the region of interest near $P_s=1$, we employ the $\lambda$ reparametrization, as depicted in Fig.~\ref{fig:probab_vs_snr}. This figure also incorporates the dependencies on the signal-to-noise ratio (SNR) of the injected GW signal and the chirp mass ${\cal M}$ of the source binary. The distributions are computed separately for three ranges of chirp mass ${\cal M}$: low, mid, and high, corresponding to ${\cal M}$ values between 13 and 17 $M_{\odot}$, 17 and 21 $M_{\odot}$, and 21 and 26 $M_{\odot}$, respectively. The histograms on the right-hand side are computed with the samples of the ${\signal}$ class that are classified with $P_s=1$, showing their distribution in terms of SNR for the three chirp mass ranges.

It is worth noting that we have either $\lambda \lesssim 7.5$ or $\lambda = +\infty$ (i.e., $P_s=1$). 
\footnote{This is because we are using single-precision float numbers so the closest $P_s$ can get to 1, without being 1, is $P_s = 1 - 2^{-24}$ and for this value of $P_s$ we have $-log_{10}(1 - P_s) = 7.22$}
In a sense, the latter case seems to accumulate all samples with $\lambda \gtrsim 7.5$. From the left column of the figure, it is apparent that the IT classifier assigns larger $\lambda$ (or $P_s$) values more uniformly over the full range of chirp mass and to lower SNR. In contrast, the CNN fails to do so for the lower chirp mass interval shown in blue. This is confirmed by the histograms in the right column, which indicate that the TCN and IT classifiers have a higher overall count (approximately 76\%, compared to 65\% for CNN) and extend to lower SNR values.

\subsection{Global assessment with Receiver Operating Characteristics}
\label{sec:ROC}

\begin{figure}
  \centering
  \includegraphics[width=\linewidth]{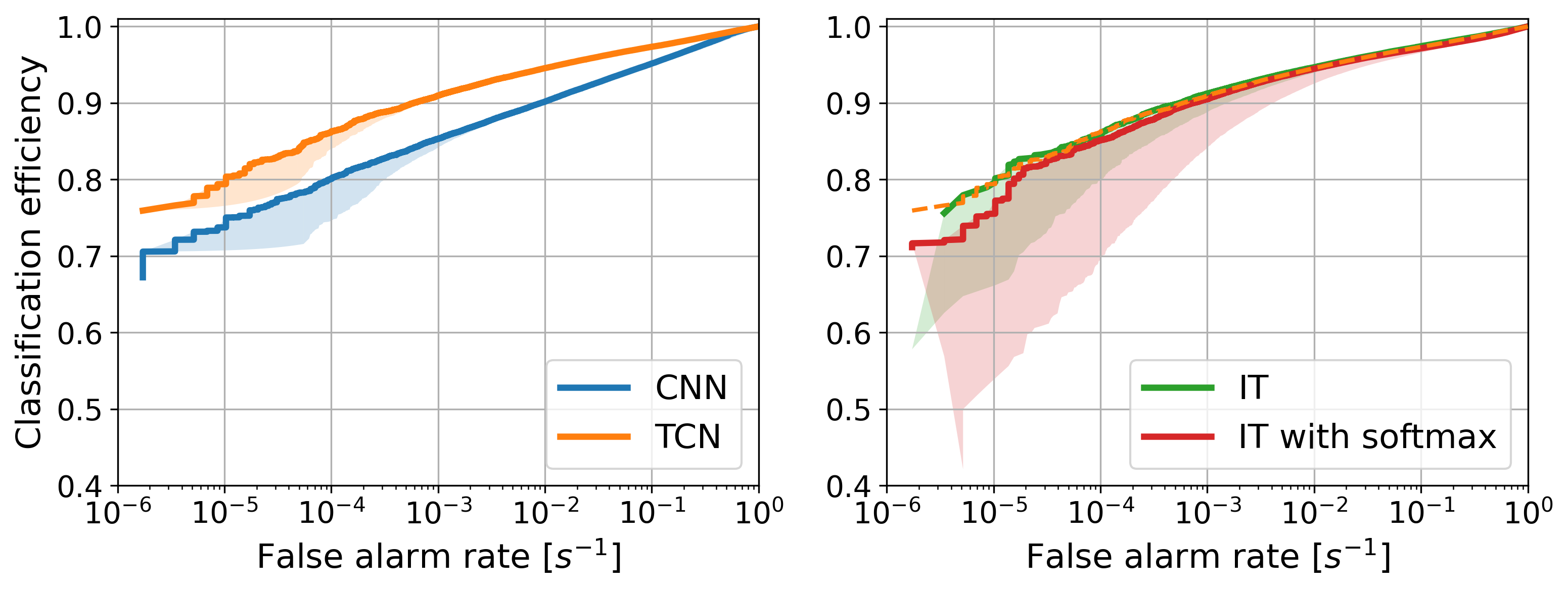}
  \caption{ROC curves for the three considered classifiers, CNN, IT, and TCN, illustrating the classification efficiency versus the false positive rate. Each classifier has been trained 10 times, and the continuous line represents the result obtained for the best model, while the shaded area covers the range from the best to the worst model. The left panel displays the TCN (orange) and CNN (blue) ROC curves. In the right panel, the ROC curves are shown for two instances of the IT architecture: one trained with softmax activation (red) and another without softmax activation (green) (refer to Sec.~\ref{sec:softmax}). The TCN ROC curve is reproduced in this panel as a dashed orange line to facilitate comparison.}
  \label{fig:roc}
\end{figure}

To fully characterize the performance of the classifier, the noise rejection and signal extraction capabilities have to be evaluated jointly. This can be done by computing the ROC curves \cite{FAWCETT2006861}. 
The classification efficiency $S_{th}/S_{tot}$ and false alarm rate $N_{th}/N_{tot}$ are evaluated from the testing set, with $S_{th}$ and $N_{th}$, the number of signal samples and noise and glitch samples with a $P_s$ value above some threshold, and $S_{tot}$ and $N_{tot}$ the total number of samples for each category. 
Note that since each sample has a duration of 1 second, $N_{tot}$ is intended here as the total duration in seconds of noise and glitch samples, so $N_{th}/N_{tot}$ is measured in $\rm{s}^{-1}$.
By varying the threshold, one obtains the ROC curves in Fig. \ref{fig:roc} which displays the classification efficiency versus the false alarm rate. The TCN and IT classifiers appear to have similar ROC curves and show a clear improvement with respect to CNN. Fig.~\ref{fig:roc} also shows the ROC computed for two instances of the IT architecture, with and without the softmax activation during training (see Sec.~\ref{sec:softmax} for a discussion).

\begin{figure}
  \centering
  \includegraphics[width=0.9\linewidth]{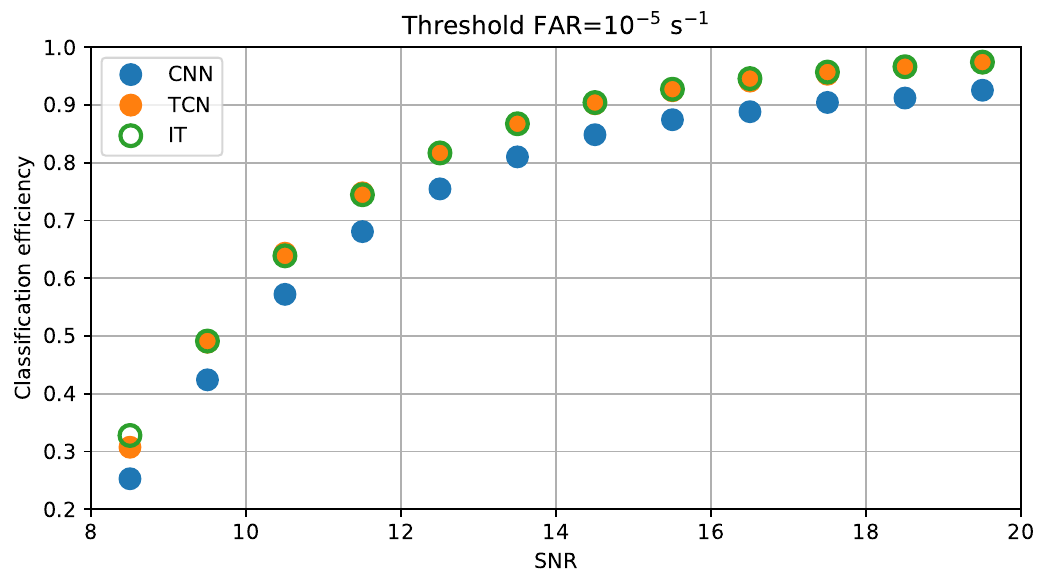}
  \caption{Classification efficiency versus SNR for a false alarm rate of $10^{-5}$ s$^{-1}$. The classifiers TCN and IT give similar efficiencies and surpass uniformly over CNN. Overall, signals with SNR=10 can be detected at the considered significance level with a good probability, larger than 50\%.}
  \label{fig:eff_snr}
\end{figure}

Fig. \ref{fig:eff_snr} shows the classification efficiency for a given false alarm rate set to $10^{-5}$ s$^{-1}$, as a function of the injected SNR as defined in Eq.~(\ref{eq:snr_opt}). The classifiers TCN and IT give similar efficiencies and surpass uniformly over CNN. Note that the efficiency shown in this figure is averaged over the full chirp mass range and thus does not show the differences evidenced in Fig.~\ref{fig:probab_vs_snr}. Overall, signals with SNR=10 can be detected at the considered significance level with a good probability, larger than 50\%.

\section{Application to the remaining O1 single-detector data}
\label{sec:remaining}

This section presents the results of applying the different classifiers to the remaining O1 data from the Livingston detector. Our primary focus is on the IT classifier, while the results for the other models can be found in \ref{app}.

\subsection{Analysis of known O1 GW events}
\label{sec:known_events}

\begin{figure}
  \centering
  \includegraphics[width=\linewidth]{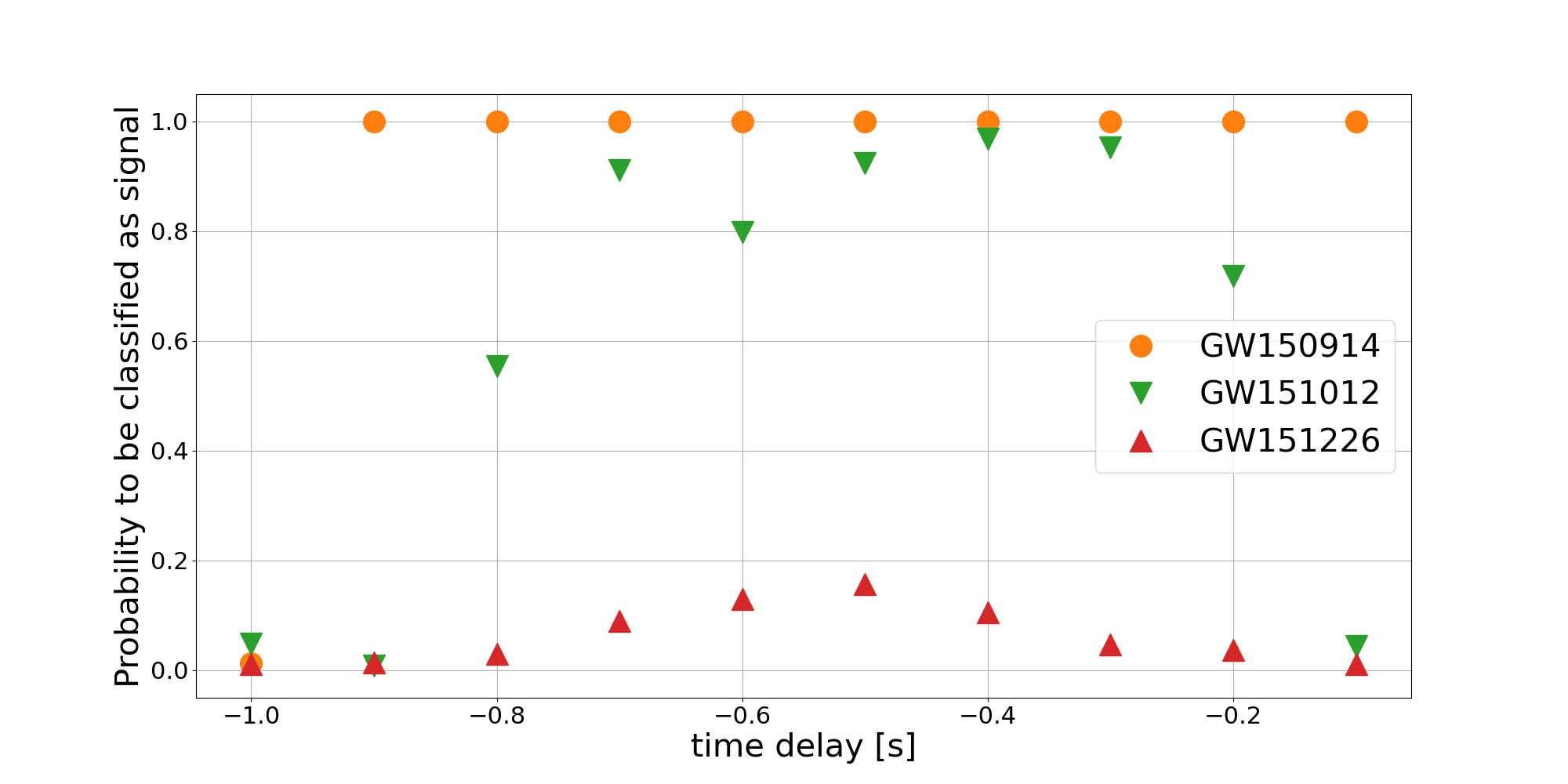}
  \caption{Evolution of the statistic $P_s$ produced with IT classifier versus the relative delay $\Delta t$ of the analysis window to the O1 event merger time (GW150914, GW151012 and GW151226). For $\Delta t = -1$ s, the analysis window only includes the initial part of the signal (inspiral), whereas, for $\Delta t = 0$ s, the analysis window starts at the merger time and thus only includes the final part (merger and ringdown).}
  \label{fig:GWeventsIT}
\end{figure}

We first investigate how the three events detected in the O1 data \cite{2019PhRvX...9c1040A} by matched filtering searches are classified by the considered models. The statistic $P_s$ is evaluated for different positions of the one-second window, that is for different time delays $\Delta t$ between the start of the analysis window and the merger time. This definition implies that, for $\Delta t = -1$ s, the analysis window only includes the initial part of the signal (inspiral), whereas, for $\Delta t = 0$ s, the analysis window starts at the merger time and thus only includes the final part (merger and ringdown). Fig.~\ref{fig:GWeventsIT} shows the evaluation of $P_s$ between those two extreme cases for the IT model for GW150914, GW151012 and GW151226 (see also \ref{app}).

As expected, when the chirp signal is not included in the analysis window, the classifier is not able to detect the presence of the signal. GW150914 appears to be loud enough to be always identified, regardless of its position in the time window, even if it is partially visible. GW151012 is only detected when the chirp is at the center of the analysis window. GW151226 is not detected. This is expected as the binary component masses are outside the range used to generate the astrophysical signals in the {\signal} class of the training data. Both events have single detector optimal SNRs for Livingston from parameter-estimation analyses lower than the minimum value of 8 we used to train the network (namely, $5.8^{+1.2}_{-1.2}$ for GW151012 and $6.9^{+1.2}_{-1.1}$ for GW151226 according to Table V of \cite{2019PhRvX...9c1040A}).

\subsection{Analysis of the remaining O1 data}
\label{sec:remaining_O1}

We analysed all the remaining L1 data in O1 excluding the month we used to train and test the classifiers (see Sec. \ref{sec:setup}). This corresponds to the period between GPS=1126051217 (2015-09-12 00:00:00 UTC) and GPS=1132444817 (2015-11-25 00:00:00 UTC) and between GPS=1135036817 (2015-12-25 00:00:00 UTC) and GPS=1137254417 (2016-01-19 16:00:00 UTC). In this period we excluded the intervals of $\pm$ 1 second around the chirp time of the 3 known events (see previous section). This amounts to a total of 4,216,489 s (about 49 days), of which 1,054,564 s (about 12 days) are single-detector times, corresponding to 25\% of the total. This data set is whitened following the same procedure used to produce the training set (the ASD was calculated from periods of non-interrupted data taking with 26 s minimum and 146,978 s maximum). The data are then divided into non-overlapping one-second segments that are processed through the three classifiers. For each, we used the best-performing model on the testing data. 
The processing time for the full data set is about 4 hours per model on NVIDIA Tesla V100S GPUs, but most of this time is taken to load the data, the extraction of the model predictions takes about 8 min for CNN, 18 min for TCN and 52 min for IT. 
No data quality information was used, so this analysis is solely based on the gravitational-wave strain data.

\begin{figure}
  \centering
  \includegraphics[width=\linewidth]{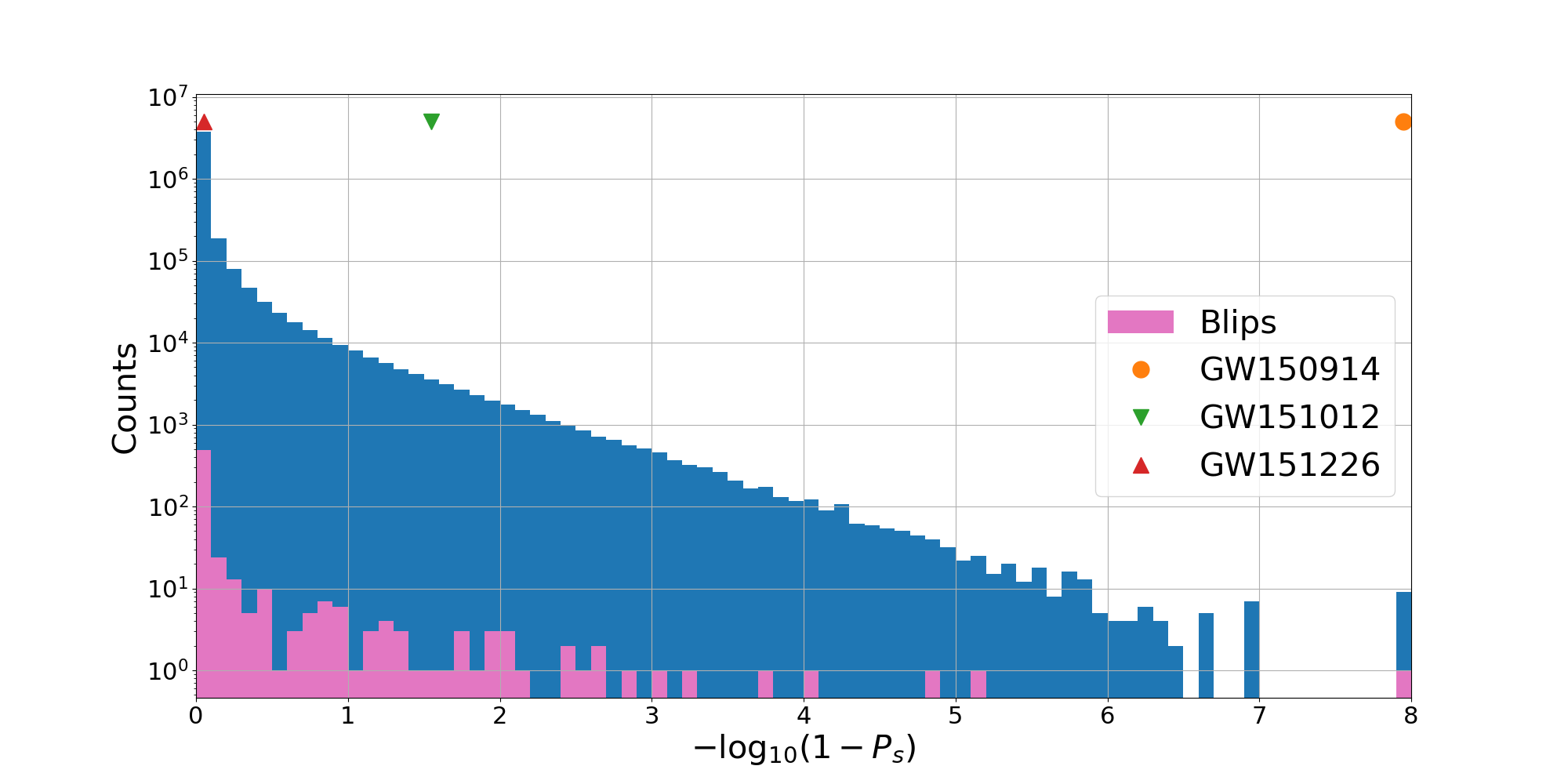}
  \caption{Distribution of the $\lambda = -\log_{10}(1 - P_s)$ statistic (shown in blue) obtained using the IT classifier on the remaining O1 dataset (refer to Sec.~\ref{sec:remaining_O1} for details). The segments with $P_s = 1$ have been assigned a value of $\lambda = 8$ for plotting purposes. The pink histogram corresponds to a subset labeled as ``Blip" glitches by {\tt Gravity Spy} \cite{glanzer_jane_2021_5649212}. The markers at the top indicate the highest values for the three O1 events displayed in Fig.~\ref{fig:GWeventsIT}. Please note that the vertical position of these markers is arbitrary.}
  \label{fig:remainingO1IT}
\end{figure}

Fig.~\ref{fig:remainingO1IT} shows the distribution of the $\lambda=- \log_{10} (1-P_s)$ statistic obtained with the IT classifier (similar plots can be found in \ref{app} for the other models). We apply the most restrictive selection cut, by requiring $P_s = 1$ (at machine precision). We recall that this selection cut corresponds to a false-alarm rate of $\lesssim 4 \times 10^{-6}$ s$^{-1}$ (that is one false alarm per 3 days) and a classification efficiency of 76\% when estimated on the testing set, see Sec.~\ref{sec:noise_reject} and \ref{sec:signal_extract}. Based on these results, we estimate from basic counting statistics that the maximum number of false alarms expected for this analysis should be $29$, $43$ and $55$ for CNN, TCN and IT respectively at 95\% level for the full data set, and $9$, $13$ and $16$ when restricting to the single-detector part. 

For the IT classifier, a total of nine segments pass the selection cut, with two occurring in single detector time at GPS=1131289775 (2015-11-11 15:09:18 UTC) and GPS=1135945474 (2016-01-04 12:24:17 UTC). For the CNN and TCN classifier, we obtain 4 and 105 segments passing the cut, with 2 and 14 falling in single detector periods. The results are thus consistent with the expectations for CNN and IT, while there is a clear excess with TCN. We have observed that a significant fraction of the triggers comes from two time intervals around 2015-10-20. Our interpretation is that the data from those periods could differ in nature from those of the training set, and TCN may be sensitive to this difference.

Interestingly there is only one segment passing the selection cut for all three classifiers: GPS=1135945474 (2016-01-04 12:24:17 UTC) which we investigate further in the next section.
As single-detector searches cannot employ statistical resampling techniques with time shifts \cite{Was:2009vh}, we can only provide an upper limit on the false alarm rate for this detection. The upper limit is estimated to be 1 event every 49 days, based on the available data from the three-month analysis period. This segment on 2016-01-04 corresponds to the event identified in the Livingston detector data during the O1 single-detector periods using a standard matched-filtering-based search, as reported in \cite{Nitz:2020naa}. However, this candidate is subsequently eliminated by the authors of Ref. \cite{Nitz:2020naa} after examining the residual obtained by subtracting the best-fit waveform from the data, since excess power is observed in the residual at frequencies below 80 Hz.

\subsection{Detailed analysis of the 2016-01-04 event}
\label{sec:20160104} 

\begin{figure}
  \centering
  \includegraphics[width=\linewidth]{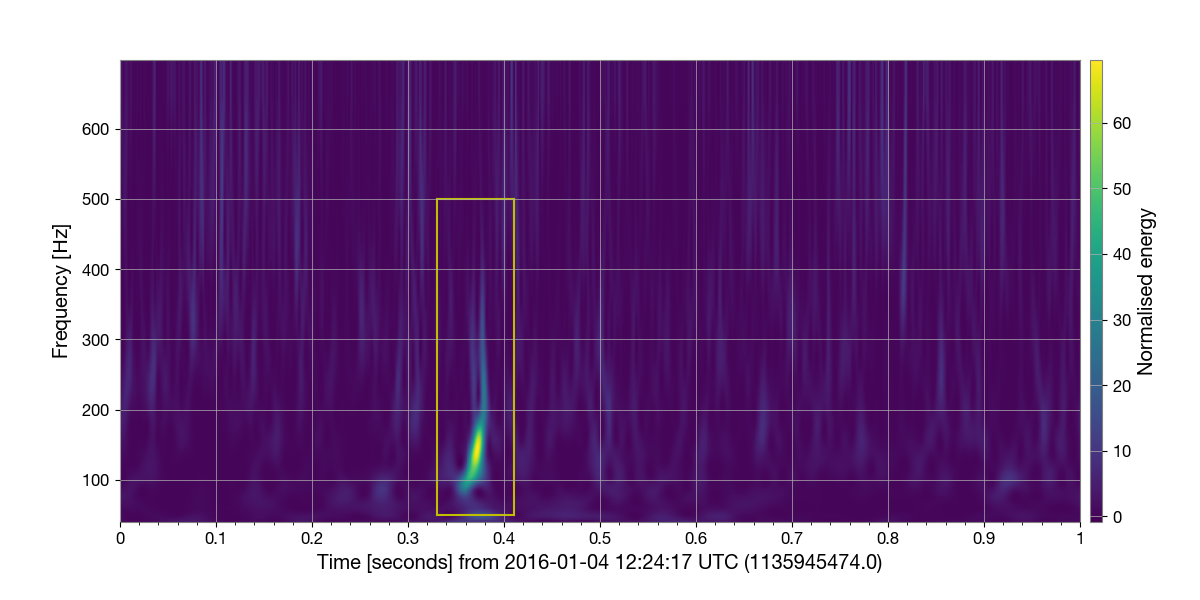}
  \includegraphics[width=\linewidth]{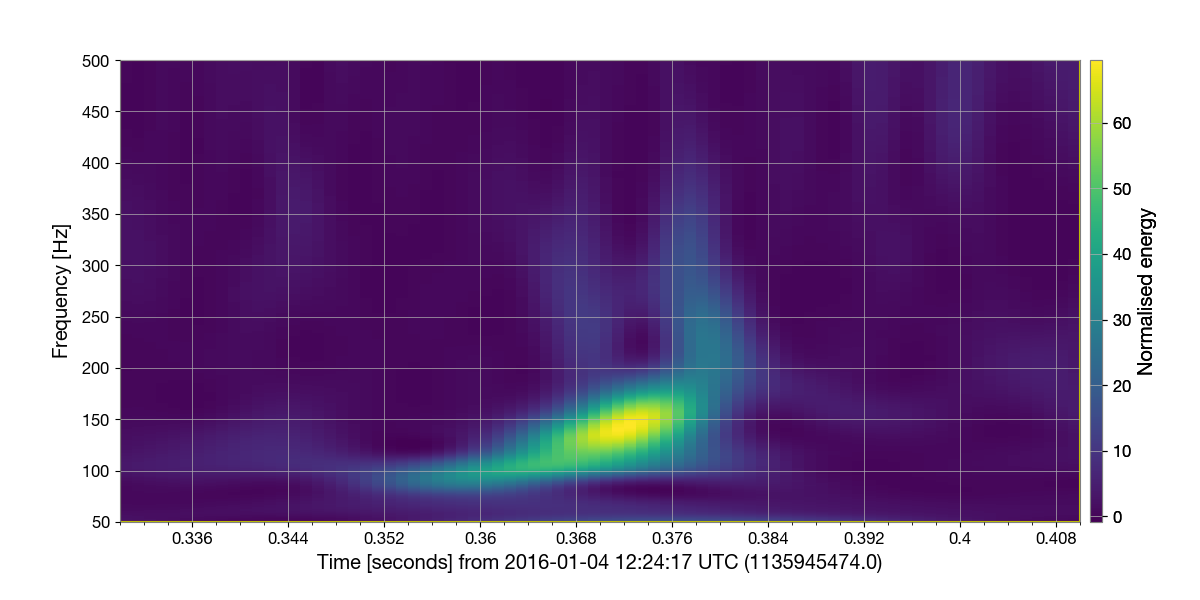}  
  \caption{Time-frequency representation of the segment at 2016-01-04 12:24:17 UTC (GPS=1135945474 s) recorded by the LIGO Livingston detector. The top panel shows the entire segment. The bottom panel is a detailed view that focuses on the transient signal at $t \sim 0.35$ and $f \sim 150$ Hz. The frequency of the signal is distinctly increasing in a chirping pattern.}
  \label{fig:q-view}
\end{figure}

We have performed a number of detailed checks of the 2016-01-04 event. We have performed a ``visual'' inspection with the time-frequency Q-transform \cite{Chatterji:2004qg}. Fig.~\ref{fig:q-view} provides a time-frequency representation of the entire segment with a Q-scan \cite{Chatterji:2004qg}. A transient is visible $\sim 0.35$ seconds after the start of the segment, at a frequency of about 150 Hz. In the magnified view, the shape of the transient is clearly indicative of a frequency modulated chirp-like transient.

The {\tt Gravity Spy} database \cite{glanzer_jane_2021_5649212} has marked this specific GPS time classified as being an instrumental artefact of the ``Blip'' type. The term refers to a well identified family of instrument glitches whose origin is still largely unknown (see, e.g., \cite{2019CQGra..36o5010C,2021CQGra..38m5014D} for more details). Generally, ``Blip'' glitches do not exhibit a chirping frequency (see Fig. 1 of \cite{Glanzer:2022avx} for a typical example). To complement this initial inspection, Fig. \ref{fig:remainingO1IT} gives in pink the statistic $\lambda$ (or equivalently $P_s$) of the 600 blip glitches listed in {\tt Gravity Spy} overlapping with the part of the O1 dataset being analyzed. The resulting distribution is compatible with the overall background distribution. The Jan 4 segment appears to be an outlier with respect to the blip glitches identified in the data.

Further, we checked if the transient signal can be fitted by a GW waveform model associated to a compact binary merger. To do so, we ran the Bayesian inference library \texttt{Bilby} \cite{Ashton_2019} and used the \texttt{IMRPhenomXPHM} waveform model \cite{PhysRevD.103.104056}. It is assumed that the component spins are co-aligned with the orbital momentum. For the rest of the source parameters, generic and agnostic priors are assumed, along with a standard $\Lambda$-CDM cosmology model with $H_0 = 67.9\,\mathrm{km}\,\mathrm{s}^{-1}\,\mathrm{Mpc}^{-1}$ \cite{Planck:2015fie}. The analysis did not include a marginalization over calibration uncertainties. The analysis results in a signal-versus-noise log Bayes factor of 47. The estimated time of arrival of the merger at the detector is GPS=$1135945474.373^{+0.076}_{-0.07}$ and the measured optimal SNR is $11.34^{+1.8}_{-1.6}$.

\begin{figure}
  \centering
  \includegraphics[trim=100 0 100 0,clip,width=\linewidth]{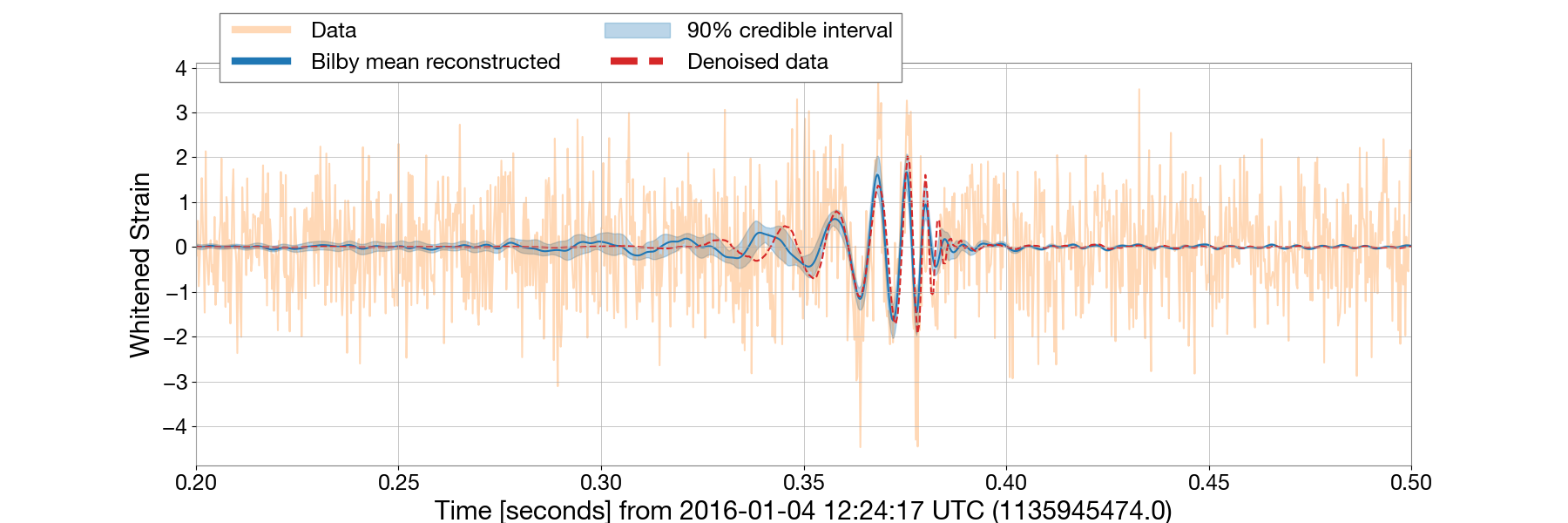}
  \caption{Comparison of the whitened L1 data (orange line) with the reconstructed waveform from the \texttt{Bilby} posteriors (blue) and the ML denoising convolutional autoencoder neural network described in \cite{denoiser} (dashed red line).}
  \label{fig:view}
\end{figure}

\begin{figure}
  \centering
  \includegraphics[width=\linewidth]{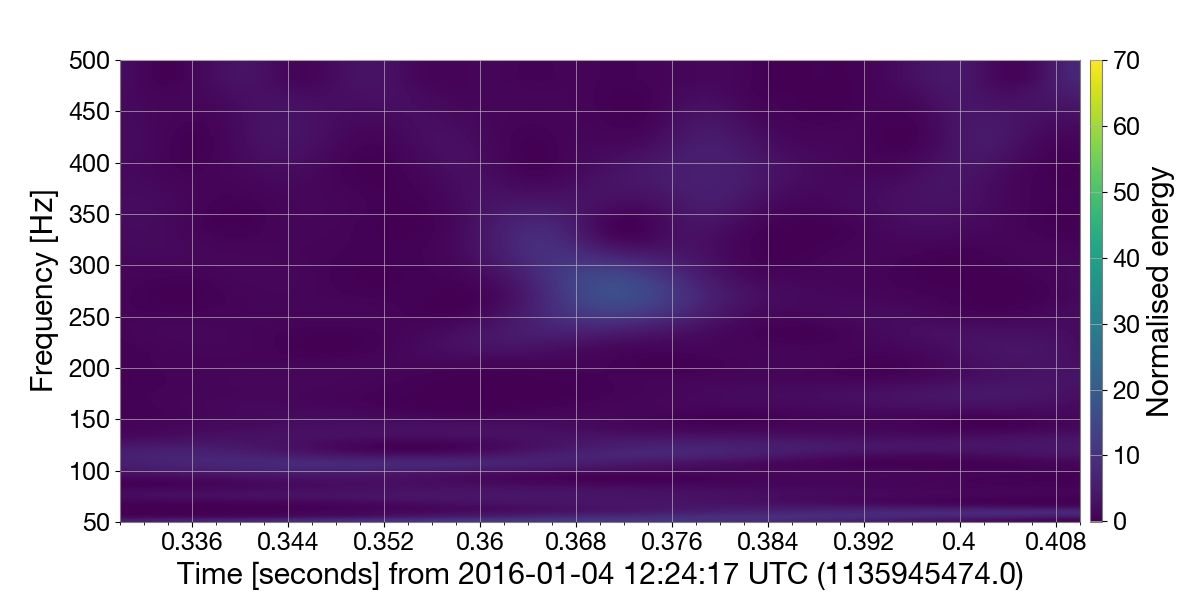}
  \caption{Time-frequency representation of the residual after the subtraction of the reconstructed waveform from \texttt{Bilby} posteriors from the data segment at 2016-01-04 12:24:17 UTC (GPS=1135945474 s). The dynamic range and color code are the same as in Fig. \ref{fig:q-view}. No excess power is visible in this plot.}
  \label{fig:qscan_res}
\end{figure}

Fig. \ref{fig:view} shows the result of the fit in the time domain, by comparing the whitened data in orange to the inferred waveform (blue) with a 90\% credible belt. 
We report that there is no significant residual after subtraction of the inferred waveform as shown in Fig. \ref{fig:qscan_res}. 
As an independent check of the nature of the signal, the figure also includes the waveform estimate produced by the denoising convolutional autoencoder described in \cite{denoiser} (dashed red). The two reconstructed waveforms are in good agreement, following a similar phase evolution, except for the initial and final parts of the signal, where the denoiser's reconstruction is not optimal because of its low-frequency cut-off, and the rather low SNR of the signal. 
In addition, we note that the denoising autoencoder was trained on the waveform family \texttt{SEOBNRv4} which is different than the one used for \texttt{Bilby} (\texttt{IMRPhenomXPHM}), which may contribute to differences.

\begin{figure}
  \centering
  \includegraphics[width=\linewidth]{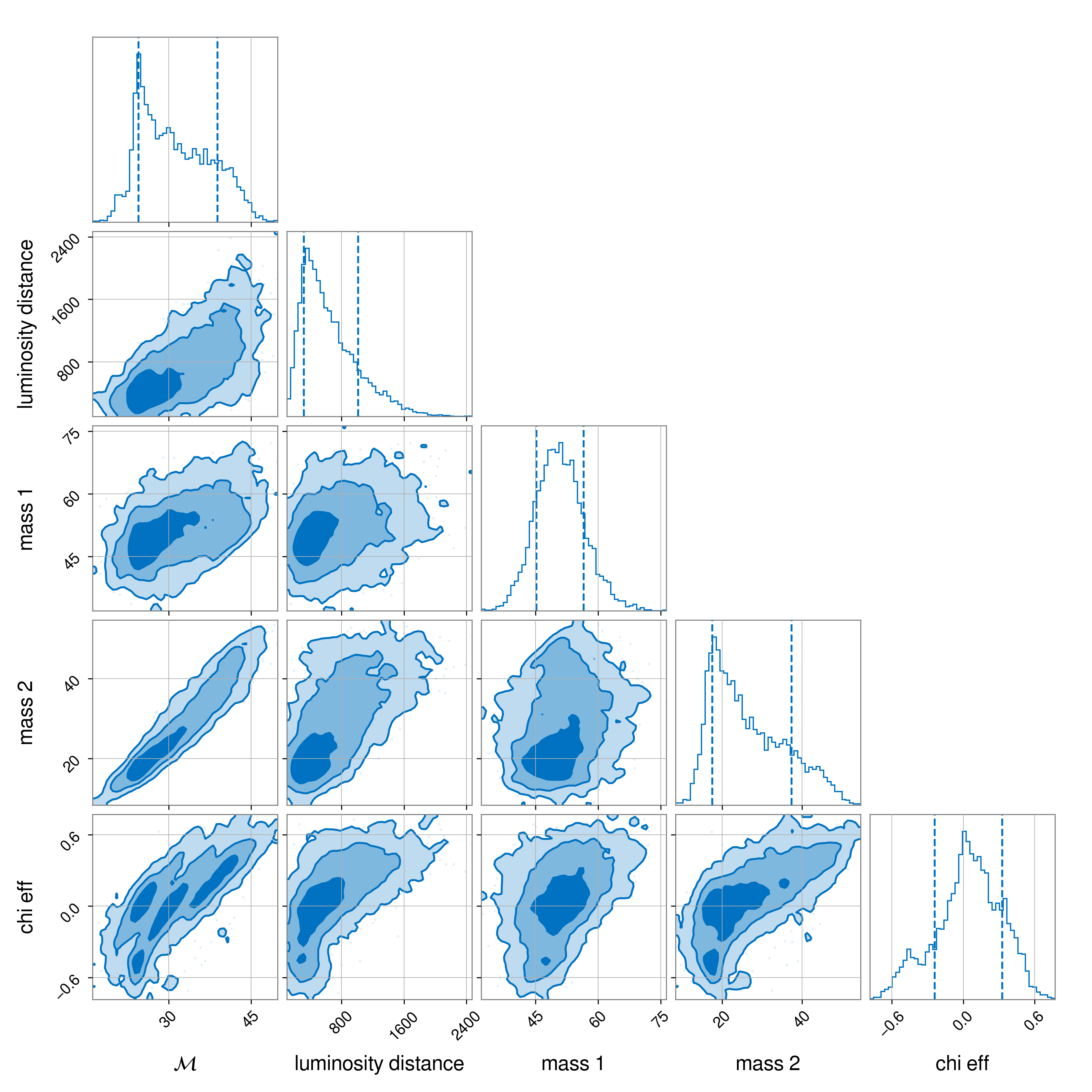}
  \caption{Posterior distribution of the chirp mass ${\cal M}$, luminosity distance, component masses $m_1$ and $m_2$, and effective spin $\chi_{\mathrm{eff}}$ for the 2016-01-04 event (see Sect.~\ref{sec:20160104} for details).} 
  \label{fig:corner_plot}
\end{figure}

Above checks are all compatible with the event being of astrophysical origin. The corner plot in Fig.~\ref{fig:corner_plot} displays the posterior distribution of the source parameters including the binary component masses, spins and source distance. Since only one detector is available, the source direction is not localized in the sky. The 90\% credible intervals for those parameters are: the measured (redshifted) chirp mass ${\cal M} = 30.18^{+12.3}_{-7.3} M_\odot$, the (redshifted) component masses $m_1 = 50.7^{+10.4}_{-8.9}\,M_{\odot}$ and $m_2 = 24.4^{+20.2}_{-9.3}\,M_{\odot}$, the binary effective spin $\chi_{\mathrm{eff}} = 0.06^{+0.4}_{-0.5}$ and the luminosity distance $d_L = 564^{+812}_{-338}\ \mathrm{Mpc}$; see \cite{Christensen:2022bxb} for a definition of those physical parameters. Overall, these values are consistent with the observed population of BBH to date.

\section{Conclusions}
\label{sec:conclusion}

This contribution demonstrates the viability of training neural network classifiers on real detectors' data for analyzing single-detector observing periods of ground-based GW detectors. We show that architectures specifically designed for time-series classification, such as IT or TCN, outperform the standard CNN typically used so far. Their relative detectability limit in terms of signal-to-noise ratio is lower by few percents to 15\% for 50 and 90\% classification efficiencies respectively. The models were trained with one month of the observing run O1 data from the LIGO Livingston detector. When applied to the remaining three months of O1 data, the classifiers independently detect a plausible GW signal of astrophysical origin on January 4, 2016. This candidate signal was also identified by \cite{Nitz:2020naa} using standard matched filtering techniques. While \cite{Nitz:2020naa} downgraded the event as a noise artifact, various diagnostics we performed substantiates the possibility of its astrophysical origin.

Operationally, we propose an approach where the multiple detector data from the first month of an observing run, labeled by standard matched filtering-based pipelines, are used to train the neural network models. The resulting classifiers can then be applied to the remaining data collected during single-detector periods. Once trained, the computational cost is such that the classifiers can produce low-latency triggers. However, the poor sky localization obtained with only one detector limits the relevance of this approach.

The current approach faces two limitations: (i) using real data for training and testing inherently limits the statistical characterization of these algorithms and their noise rejection capabilities, as already highlighted in \cite{Schafer:2022dxv} and observed with the excess of triggers produced by the TCN classifier; (ii) there is a technical issue arising from the use of bounded selection statistics (i.e., class membership probabilities in our case) that leads to numerical intricacies. More generally, due to the absence of a mathematical theory for neural networks, their precise statistical characterization on noisy data remains an open question. Consequently, research in this field is limited to a trial and error heuristic approach.

This contribution opens up new possibilities for analyzing the fairly large single-detector data set. Applying the proposed classifiers to other LIGO-Virgo observing runs and broadening the parameter space to include lower masses and effects such as higher-order modes or precession would be interesting directions for future work.

\section{Acknowledgements}

This work was partially supported by European Union’s Horizon 2020 research and innovation programme under grant agreement No 653477, diiP (data intelligence institute of Paris), IdEx Université de Paris, ANR-18-IDEX-0001, and the COST action G2net (CA 17137). MB gratefully acknowledges partial support from the Polish National Science Centre grants no. 2016/22/E/ST9/00037 and 2021/43/B/ST9/01714, and Poland's high-performance Infrastructure PLGrid (ACK Cyfronet AGH). The authors are grateful for computational resources provided by the LIGO Laboratory and supported by National Science Foundation Grants PHY-0757058 and PHY-0823459. This work was granted access to the HPC resources of IDRIS under the allocation 2021-A0111012956 and 2021-AD011012279 made by GENCI. Some of the numerical computations were performed on the DANTE platform, APC, France.

The authors would like to thank Charlotte Pelletier for suggesting the investigation of the TCN and Inception Time classifiers. The authors also thank Liliia Sinitsyna, Anirudh Kalla, Hugo Marchand and Félix Bretaudeau that have contributed to this project during their internship. É CM would like to thank Cecilio Garcia-Quiros for useful advice with the usage of \texttt{Bilby}, and Konstantin Leyde for stimulating discussions about this work. We thank Sophie Bini and Thomas Dent for their comments during the LVK internal review.

This research has made use of data or software obtained from the Gravitational Wave Open Science Center (\url{gwosc.org}), a service of the LIGO Scientific Collaboration, the Virgo Collaboration, and KAGRA. This material is based upon work supported by NSF's LIGO Laboratory which is a major facility fully funded by the National Science Foundation, as well as the Science and Technology Facilities Council (STFC) of the United Kingdom, the Max-Planck-Society (MPS), and the State of Niedersachsen/Germany for support of the construction of Advanced LIGO and construction and operation of the GEO600 detector. Additional support for Advanced LIGO was provided by the Australian Research Council. Virgo is funded, through the European Gravitational Observatory (EGO), by the French Centre National de Recherche Scientifique (CNRS), the Italian Istituto Nazionale di Fisica Nucleare (INFN) and the Dutch Nikhef, with contributions by institutions from Belgium, Germany, Greece, Hungary, Ireland, Japan, Monaco, Poland, Portugal, Spain. KAGRA is supported by Ministry of Education, Culture, Sports, Science and Technology (MEXT), Japan Society for the Promotion of Science (JSPS) in Japan; National Research Foundation (NRF) and Ministry of Science and ICT (MSIT) in Korea; Academia Sinica (AS) and National Science and Technology Council (NSTC) in Taiwan.

\bibliographystyle{iopart-num}
\bibliography{main}

\appendix

\section{Additional results}
\label{app}

In addition to the figures presented with the IT classifier in Section \ref{sec:remaining}, we provide here the corresponding figures for the CNN and TCN models.

This includes the analysis conducted with known O1 GW events in Sec.~\ref{sec:known_events}. Comparing Figs.~\ref{fig:GWeventsIT} and \ref{fig:GWeventsCNN_TCN}, we observe that GW150914 is the only event classified with $P_s = 1$ by all classifiers, while GW151012 and GW151226 never satisfy this selection criterion. For TCN, the $P_s$ statistic is particularly low for both of these events, whereas CNN yields the highest $P_s$ value.

\begin{figure}
  \centering
  \includegraphics[width=\linewidth]{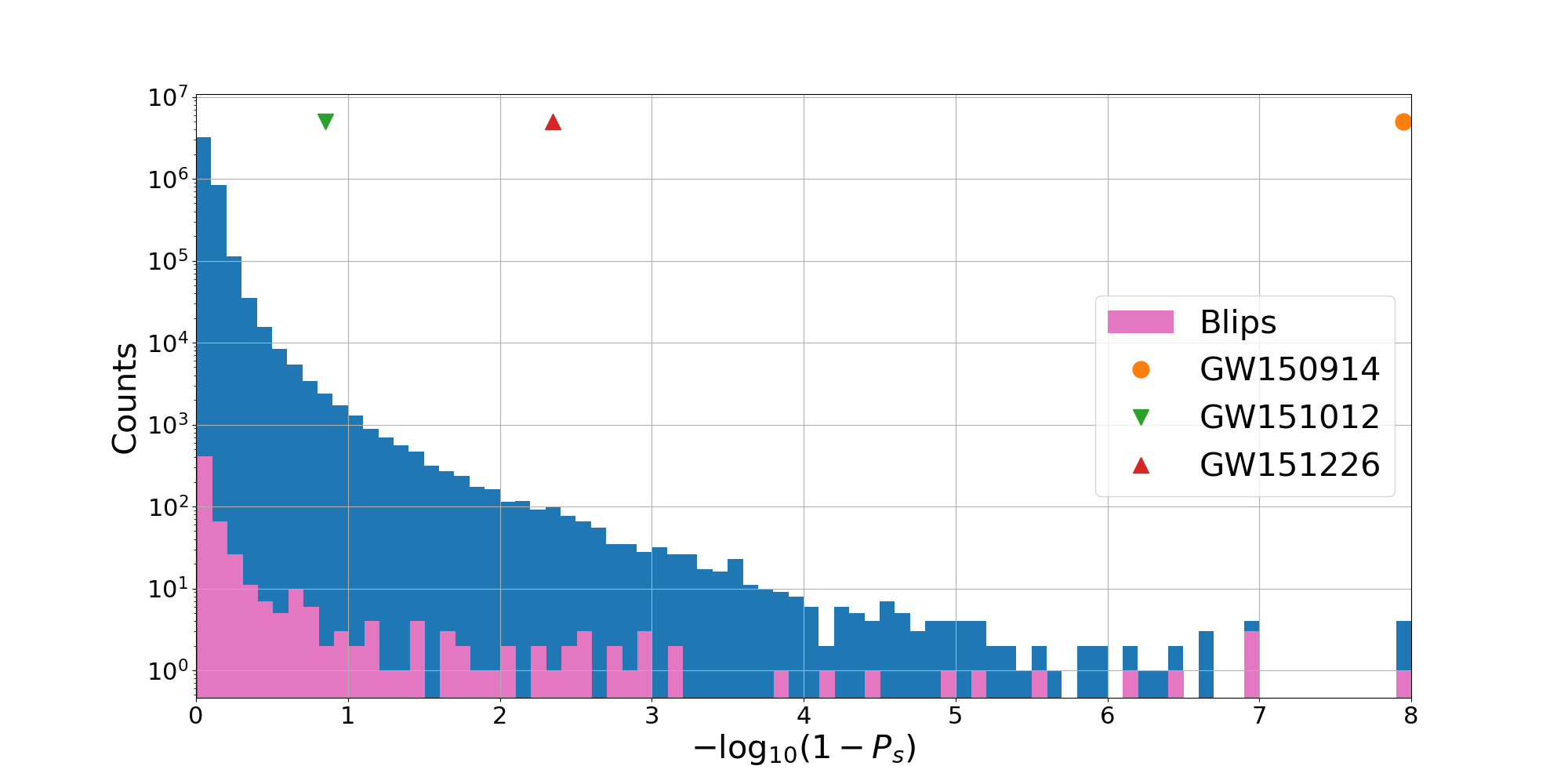}
  \includegraphics[width=\linewidth]{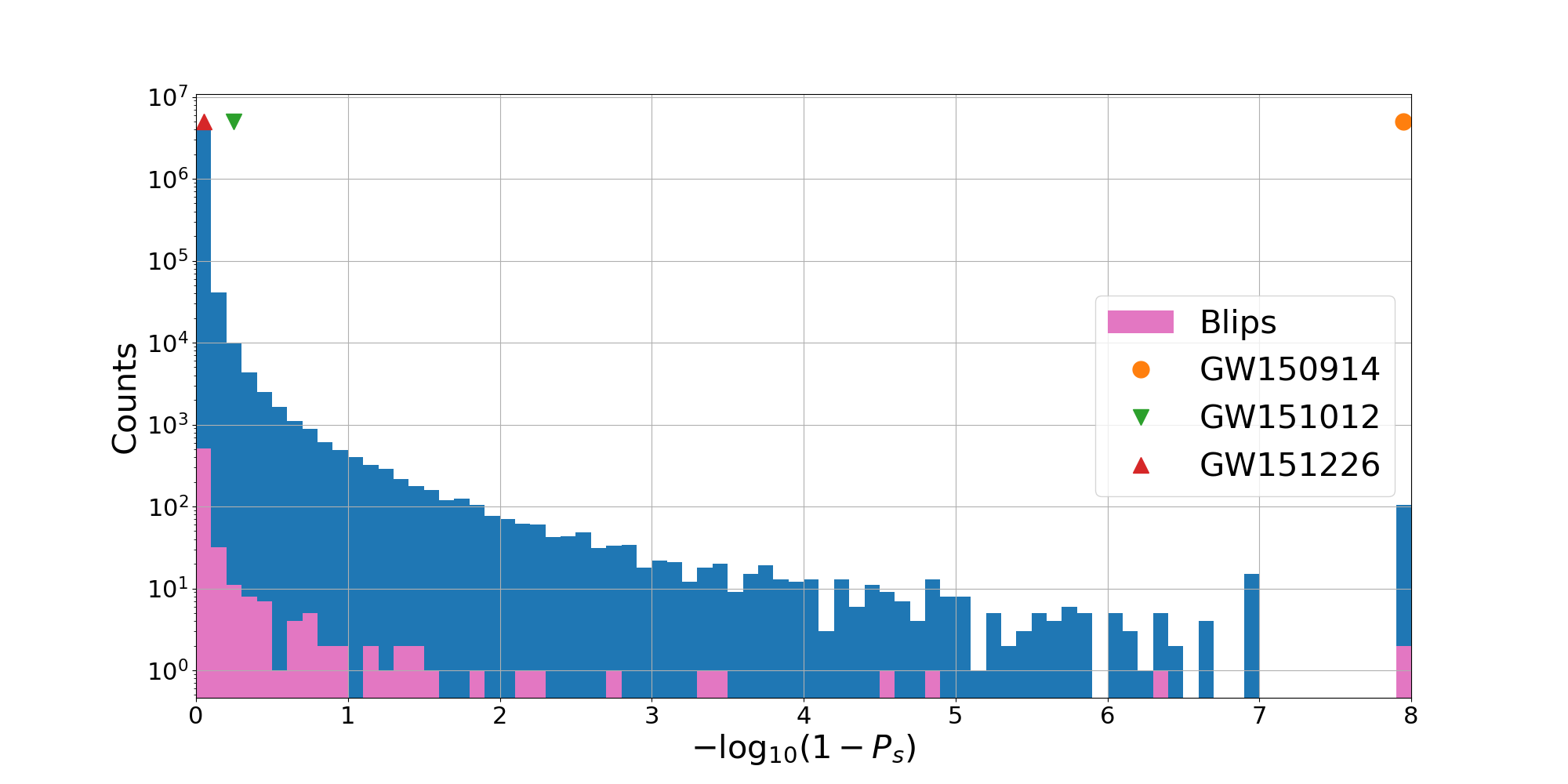}
  \caption{Distribution of the $\lambda = -\log_{10}(1 - P_s)$ statistic (shown in blue) obtained using the CNN (top panel) and TCN (bottom panel) classifiers on the remaining O1 dataset (refer to Sec.~\ref{sec:remaining_O1} for details). The segments with $P_s = 1$ have been assigned a value of $\lambda = 8$ for plotting purposes. The pink histogram corresponds to a subset labeled as "Blip" glitches by {\tt Gravity Spy} \cite{glanzer_jane_2021_5649212}. The markers at the top indicate the highest values for the three O1 events displayed in Fig.~\ref{fig:GWeventsIT}. Please note that the vertical position of these markers is arbitrary.}
  \label{fig:remainingO1CNN_TCN}
\end{figure}

\begin{figure}
  \centering
  \includegraphics[width=\linewidth]{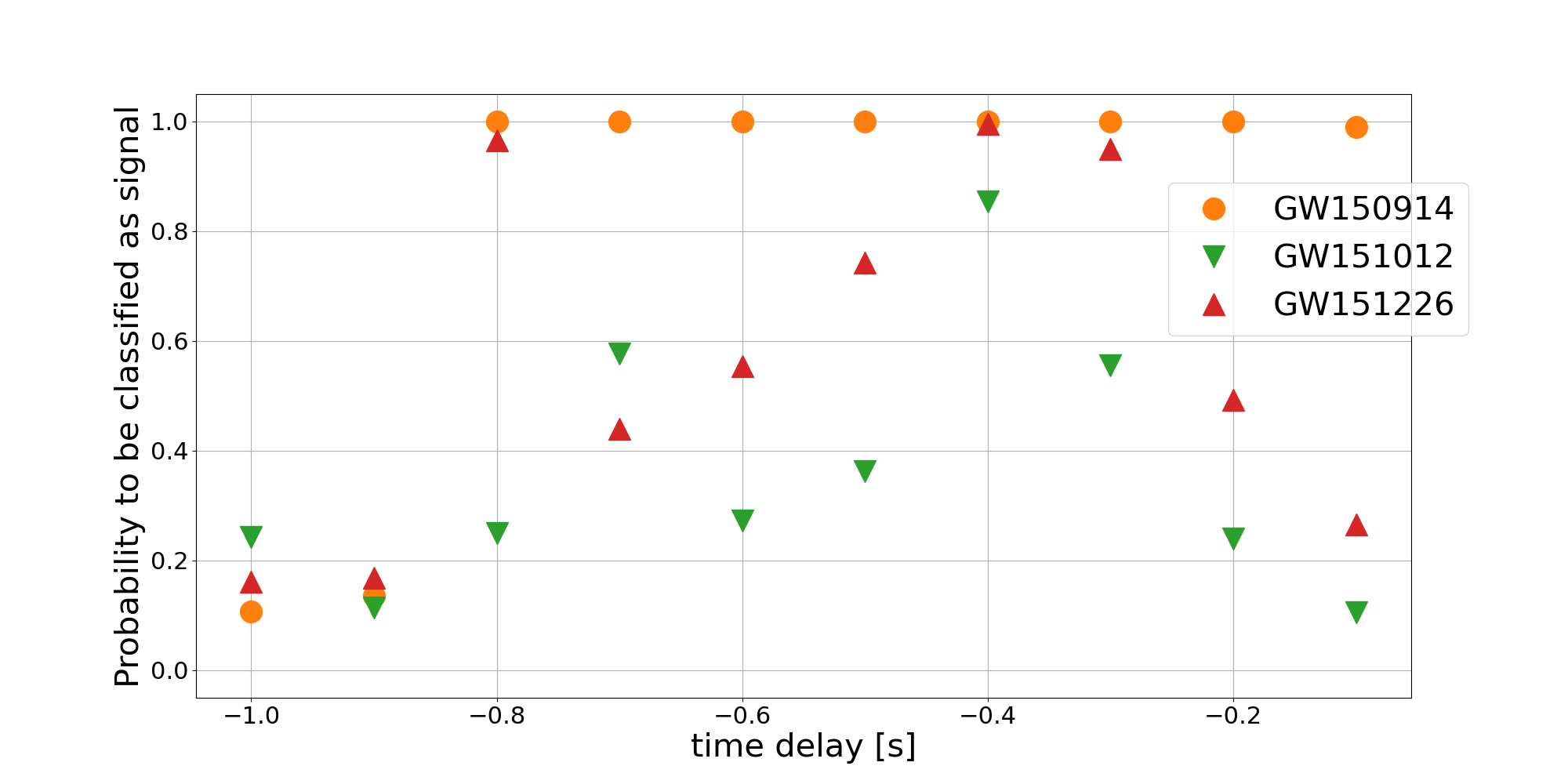}
  \includegraphics[width=\linewidth]{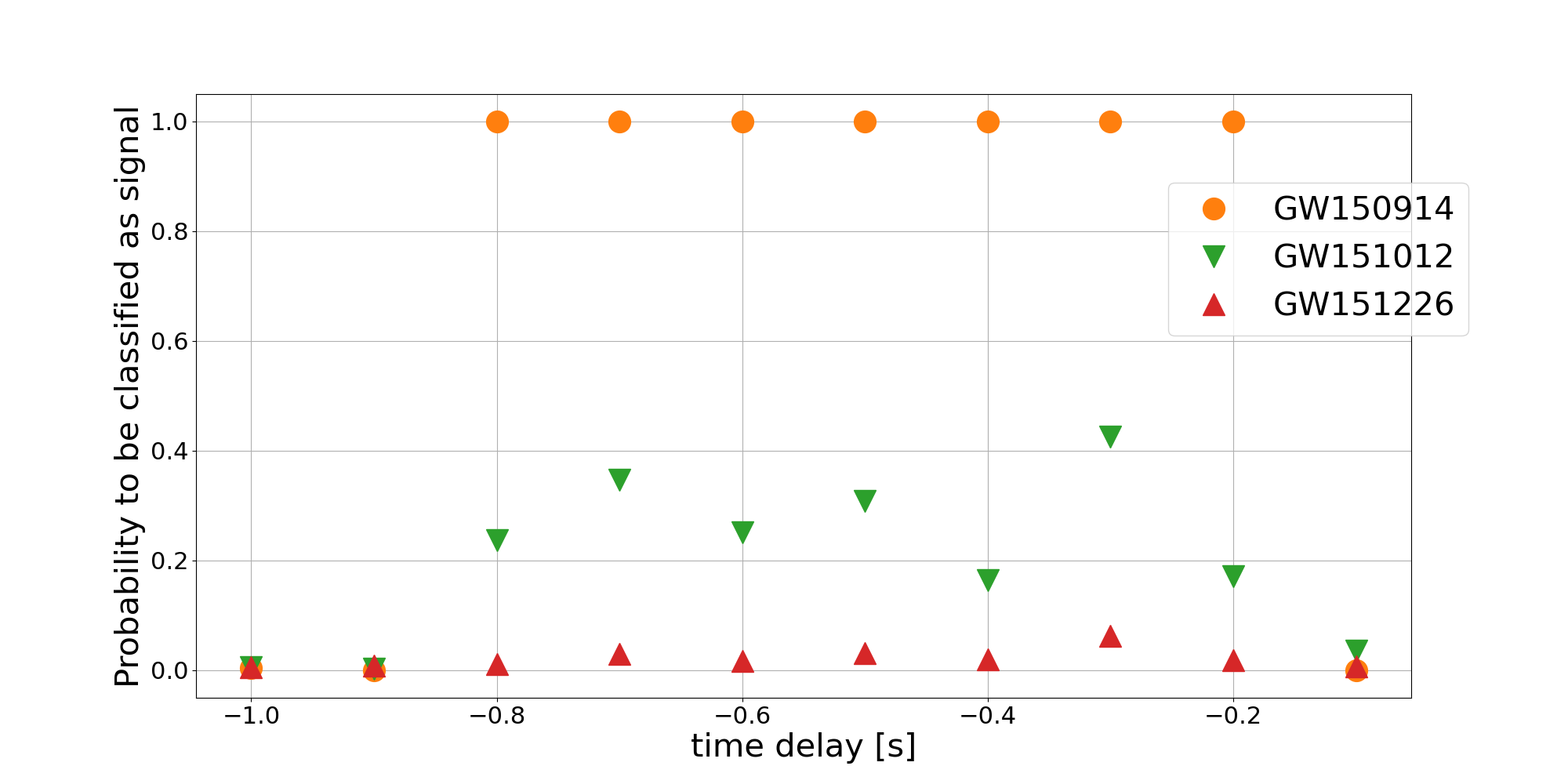}
  \caption{Top panel: evolution of the statistic $P_s$ produced with CNN classifier versus the relative delay $\Delta t$ of the analysis window to the O1 event merger time (GW150914, GW151012 and GW151226). For $\Delta t = -1$ s, the analysis window only includes the initial part of the signal (inspiral), whereas, for $\Delta t = 0$ s, the analysis window starts at the merger time and thus only includes the final part (merger and ringdown). Bottom panel: the TCN classifier results.}
  \label{fig:GWeventsCNN_TCN}
\end{figure}

We also present background histograms obtained with the remaining O1 data, similar to Fig.~\ref{fig:remainingO1IT} in Sec.~\ref{sec:remaining_O1}. Fig.~\ref{fig:remainingO1CNN_TCN} shows the same distribution for CNN and TCN. 
The distribution obtained with CNN decays faster than the other two models but exhibits a tail that reaches the extreme point, $P_s=1$. CNN appears to be more sensitive to the presence of blip glitches, as the total number of blip glitches with $\lambda > 3$ is twice as high as the number in the other two models.

\end{document}